\documentclass[]{iac}
\usepackage[acronym]{glossaries}
\usepackage{hyperref}
\usepackage[citestyle=numeric,bibstyle=numeric,maxbibnames=1,maxcitenames=1,sorting=none,sortcites]{biblatex}
\usepackage[dvipsnames]{xcolor}
\usepackage[inline]{enumitem}
\usepackage[binary-units]{siunitx}
\usepackage[nameinlink]{cleveref}
\usepackage{tabularx}
\usepackage{acro}
\usepackage[british]{babel}
\usepackage{xcolor-material}
\usepackage{soul}

\makeglossaries

\DeclareAcronym{MBSE}{
	short = MBSE,
    long = Model-Based Systems Engineering
}

\DeclareAcronym{CI}{
	short = CI,
    long = Continuous Integration
}

\DeclareAcronym{CD}{
	short = CD,
    long = Continuous Delivery
}

\DeclareAcronym{TDD}{
	short = TDD,
    long = Test-Driven Development
}

\DeclareAcronym{PCB}{
	short = PCB,
    long = Printed Circuit Board
}

\DeclareAcronym{COTS}{
	short = COTS,
    long = Commercial Off-The-Shelf
}

\DeclareAcronym{RF}{
	short = RF,
    long = Radio Frequency
}

\DeclareAcronym{UL} {
	short = UL,
    long = University of Luxembourg
}

\DeclareAcronym{SoC} {
	short = SoC,
    long = System on Chip,
}

\DeclareAcronym{IC} {
	short = IC,
    long = Integrated Circuit
}

\usepackage[many]{tcolorbox}
\newtcbox{\todo}{
    on line,
    colback=red!5!white,
    colframe=red!75!black,
    coltitle=red!75!black,
    fonttitle=\bfseries,
    fontupper=\footnotesize,
    title=TODO,
    detach title,
    before upper={\tcbtitle\ },
    nobeforeafter,
    tcbox raise base,
    top=0pt,bottom=0pt,left=0mm,right=0mm,
    toprule=0mm,
    bottomrule=0mm,boxsep=0.7mm,
}

\def\todo#1{}

\DeclareSourcemap{
  \maps[datatype=bibtex]{
    \map[overwrite]{
      \step[fieldsource=doi, final]
      \step[fieldset=url, null]
      \step[fieldset=eprint, null]
    }  
  }
}

\newbibmacro{string+url}[1]{%
    \iffieldundef{url}{#1}{\href{\thefield{url}}{#1}}
}
\DeclareFieldFormat{title}{\usebibmacro{string+url}{\mkbibemph{#1}}}
\DeclareFieldFormat[article]{title}{\usebibmacro{string+url}{\mkbibquote{#1}}}
\DeclareFieldFormat[inproceedings]{title}{\usebibmacro{string+url}{\mkbibquote{#1}}}
\DeclareFieldFormat[thesis]{title}{\usebibmacro{string+url}{\mkbibquote{#1}}}

\newcommand\myshade{85}
\colorlet{mylinkcolor}{violet}
\colorlet{mycitecolor}{Turquoise}
\colorlet{myurlcolor}{Blue}

\hypersetup{
  linkcolor  = mylinkcolor!\myshade!black,
  citecolor  = mycitecolor!\myshade!black,
  urlcolor   = myurlcolor!\myshade!black,
  colorlinks = true,
}

\usepackage{array}
\newcolumntype{L}[1]{>{\raggedright\let\newline\\\arraybackslash\hspace{0pt}}m{#1}}
\newcolumntype{C}[1]{>{\centering\let\newline\\\arraybackslash\hspace{0pt}}m{#1}}
\newcolumntype{R}[1]{>{\raggedleft\let\newline\\\arraybackslash\hspace{0pt}}m{#1}}

\setlist{leftmargin=0.8cm}

\begin{document}

\lhead{}
\chead{}
\rhead{}

\makeatletter
\lfoot{}\cfoot{}\rfoot{Page \thepage\ of \pageref{LastPage}}%
\makeatother


\title{Agile Systems Engineering for sub-CubeSat scale spacecraft}

\IACauthor{Konstantinos~Kanavouras}{University of Luxembourg, \url{konstantinos.kanavouras@uni.lu}}
\IACauthor{Andreas~M.~Hein}{University of Luxembourg, \url{andreas.hein@uni.lu}}
\IACauthor{Maanasa~Sachidanand}{ISAE-SUPAERO, France, \url{maanasa.sachidanand@student.isae-supaero.fr}}

\abstract{Space systems miniaturization has been increasingly popular for the past decades, with over 1600
CubeSats and 300 sub-CubeSat sized spacecraft estimated to have been launched since 1998. This trend
towards decreasing size enables the execution of unprecedented missions in terms of quantity, cost and
development time, allowing for massively distributed satellite networks, and rapid prototyping of space
equipment. Pocket-sized spacecraft can be designed in-house in less than a year and can reach weights
of less than 10g, taking away the considerable costs and requirements typically associated with orbital
flight. However, while Systems Engineering methodologies have been proposed for missions down to
CubeSat size, there is still a gap regarding design approaches for picosatellites and smaller spacecraft,
which can exploit their potential for iterative and accelerated development. In this paper, we propose a
Systems Engineering methodology that abstains from the classic waterfall-like approach in favor of agile
practices, focusing on available capabilities, delivery of features and design ``sprints''. 
Shifting away from the typical design-verify-operate model, this method originates from the software engineering discipline, focusing more on short design iterations, team collaboration, and focusing on quickly delivering a minimum viable product.
This allows quick adaptation to imposed
constraints, changes to requirements and unexpected events (e.g.~chip shortages or delays), by making
the design flexible to well-defined modifications. Two femtosatellite missions, currently under development
and due to be launched in 2023, are used as case studies for our approach, showing how miniature
spacecraft can be designed, developed and qualified from scratch in 6 months or less. Both missions
involve the attachment of a chip-sized satellite (“ChipSat”) into a larger spacecraft, either relying on
their host for communications and power or being completely independent.
We claim that the proposed method can simultaneously increase confidence in the design and decrease turnaround time for extremely small satellites, allowing novel and unprecedented missions to take shape without the overhead traditionally associated with sending cutting-edge hardware to space.
}
\IACkeywords{systems engineering}{femtosatellites}{attosatellites}{chipsat}{agile}{scrum}

\maketitle

\acuseall
\printacronyms[template=supertabular]



\section{Introduction}

    CubeSats are a class of spacecraft categorised as \emph{nano-satellites} \autocite{kulu_nanosats_2021}, which typically weigh less than \SI{10}{\kilo\gram} and considerably reduce the effort required to reach orbit.
    Since the publication of the CubeSat standard \cite{CDS14}, more than 1600 CubeSats have been launched into orbit \autocite{kulu_nanosats_2021}.

 
    
    While CubeSat trends already had a momentous impact in space engineering and are considered of low relative cost and effort in the space industry \autocite{poghosyan_cubesat_2017, sweeting_modern_2018}, they still require significant investment from designers and builders. Newcomers in the CubeSat world, such as educational institutions or start-ups, are usually faced with a build and launch cost of more than \SI{200000}{\$}, and at least some years of development (\Cref{tab:compi}). The effort increases further if multiple CubeSat launches are desired.

   \begin{table*}[]
    \centering
    \caption{Comparison between classes of nano-satellites and smaller \autocite{hein_attosats_2019, PQ1, speretta_cubesats_2016, saeed_cubesat_2020}}
    \label{tab:compi}
    \begin{tabular}{@{}llll@{}}
    \toprule
     & CubeSats & PocketQubes & ChipSats \\ \midrule
     & \multicolumn{1}{c}{\includegraphics[height=1.5cm]{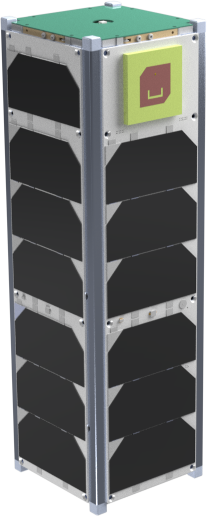}} & \multicolumn{1}{c}{\includegraphics[height=1.5cm]{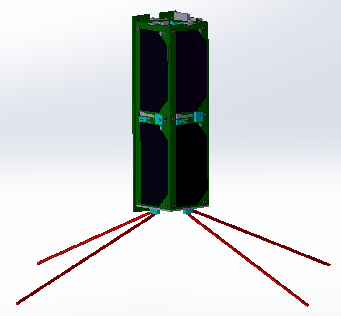}} & \multicolumn{1}{c}{\includegraphics[height=1.5cm]{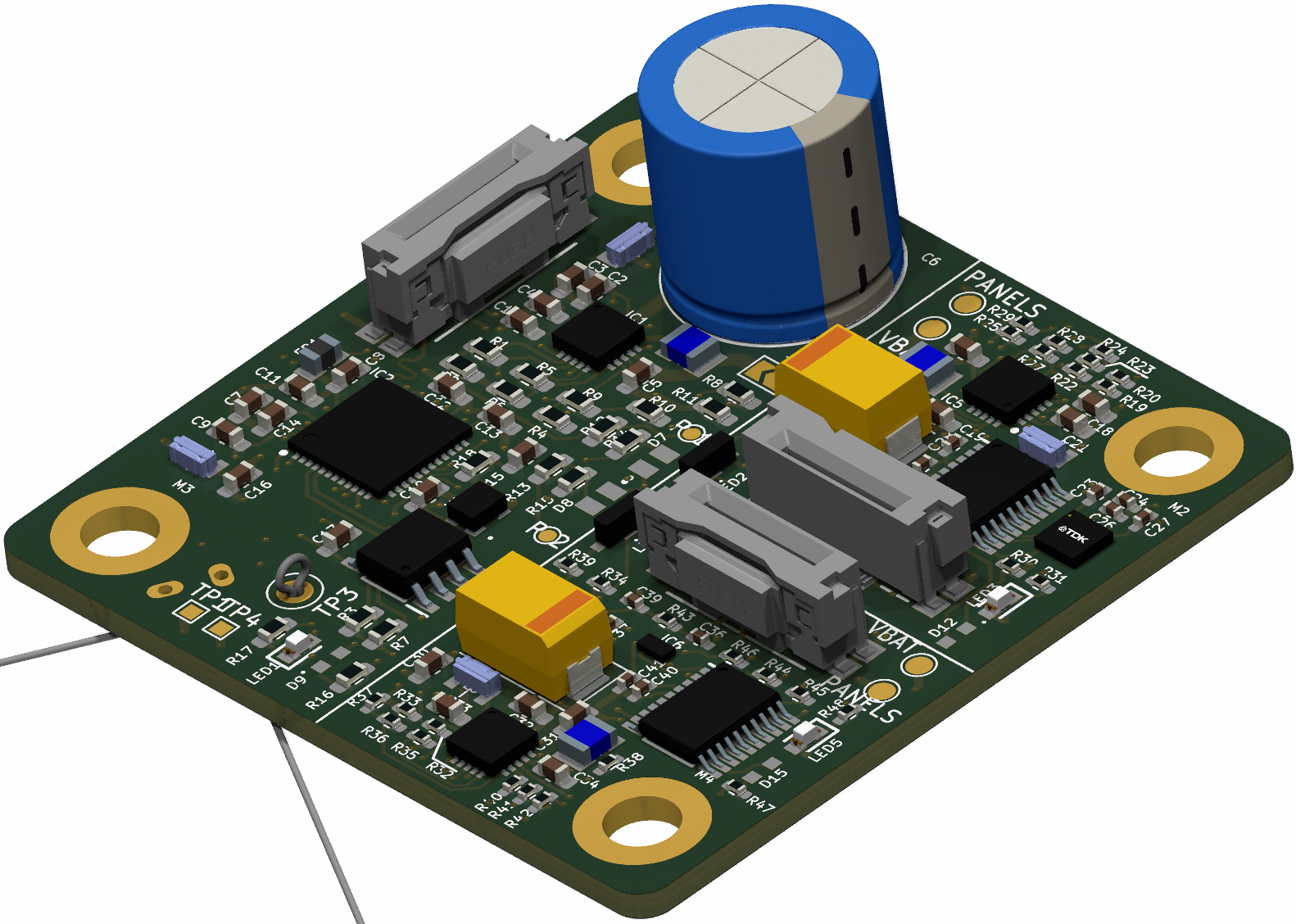}} \\
     \cmidrule(lr){2-4}
    \textbf{Mass} & \SIrange{1}{10}{\kg} & \SIrange{100}{1000}{\gram} & \SIrange{1}{100}{\gram} \\
    \textbf{Manufacturing Cost} & \SI{e5}{\$}  & \SI{e4}{\$} & \SI{e3}{\$}  \\
    \textbf{Launch Cost} & \SI{e5}{\$}  & \SI{e4}{\$} & \SI{e3}{\$}  \\
    \textbf{Development Time} & \SIrange{3}{5}{years}  & \SIrange{3}{5}{years}  & \SIrange{4}{8}{months} \\
    \addlinespace[0.5em]
    \textbf{Power} & \SIrange{1}{10}{\watt}  & \SIrange{0.5}{2}{\watt}  & \SIrange{0.1}{0.5}{\watt}  \\ 
    \textbf{RF Data Rate} & \SIrange{0.1}{400}{Mbps}  & \SIrange{10}{250}{kbps}  & \SIrange{10}{2000}{bps}  \\ 
    \bottomrule
    \end{tabular}
    \end{table*}
    
    The considerable cost and effort to create CubeSats has therefore sparked ideas for even smaller classes of spacecraft: ThinSats \autocite{twiggs_thinsat_2018},
    PocketQubes \autocite{PQ1},
    ChipSats \autocite{manchester_centimeter-scale_2015}
    and others.
    These so-called \textbf{``sub-CubeSat'' spacecraft} are often categorised, according to their mass, into: \autocite{sweeting_modern_2018,hein_attosats_2019}
    
    \begin{itemize}[itemsep=0pt]
        \item \textbf{pico-satellites} (\SIrange{100}{1000}{\gram}),
        \item \textbf{femto-satellites} (\SIrange{10}{100}{\gram}), and
        \item \textbf{atto-satellites} (\SIrange{1}{10}{\gram}).
    \end{itemize}
    
    Such smaller spacecraft take advantage of the increasing miniaturisation of off-the-shelf mechanisms and electronics to deliver value by significantly decreasing cost and development effort. The cost of a single flight model for an atto-satellite may be less than \SI{1000}{\$} \autocite{manchester_kicksat_2013}, while the development time for one unit can be less than 6 months, even with all subsystems being built in-house \autocite{hein_attosats_2019}.

    As of 2022, more than 350 known pico, femto and atto-satellites have been launched as individual units or as parts of constellations (\Cref{tab:launch}). Developers have extrapolated the CubeSat standard into 1/2U, 1/3U and 1/4U form factors \autocite{gangestad_flight_2015, kulu_nanosats_2021}. ``PocketQubes'' \autocite{PQ1} are a standardised version of modular picosatellites, which can weigh up to \SI{250}{\gram} per cubical unit, and have already started having commercial implementations \autocite{noauthor_unicorn_nodate,noauthor_fossa_2021}. Other design classes for sub-CubeSat spacecraft have been proposed, such as SunCubes \autocite{a_s_u_news_suncube_2016}, PCBSats \autocite{barnhart_enabling_2007,gong_design_2022}, and more \autocite{perez_survey_2016}.

    One of the most representative examples of sub-CubeSat missions is the KickSat mission designed in Cornell University \autocite{manchester_centimeter-scale_2015,manchester_kicksat_2013}. This mission deployed 105 satellites which belong in the "ChipSat" spacecraft class. ChipSats are atto-satellites where all components and subsystems are integrated on a single PCB board, resulting in extremely low mass and size.

    \begin{table}[bth]
    \caption{Number of known nanosatellite (and smaller) launches \autocite{kulu_nanosats_2021,abate_inexpensive_2019}}
    \label{tab:launch}
    \centering
    \begin{tabular}{@{}ll@{}}
    \toprule
    CubeSats                        & 1604 \\ 
    Nanosatellites (non-CubeSat)    & 86   \\[1ex]
    PocketQubes                     & 46   \\
    Picosatellites (non-PocketQube) & 217  \\[1ex]
    Femtosatellites                 & 4  \\
    Attosatellites                  & 105  \\ \bottomrule
    \end{tabular}
    \end{table}

    \subsection{Space Systems Lifecycle Models}
    \paragraph{}
    Space systems development has traditionally followed a top-down, waterfall-like approach \autocite{shea_nasa_2017, ECSS-E-ST-10C}. The life cycle of a project typically follows a set of predefined phases, starting from the mission conceptualisation, proceeding with the detailed design definition, and finishing with qualification and then flight \autocite{aguirre_introduction_2013}. This is often modelled as a V-diagram, showing how the system concept influences component design, leading back to system validation in the end \autocite{clark_system_2009,bundesrepublik_deutschland_v-modell_2006}. Development traditionally follows a \textit{stage-gate} process, where a project is evaluated and its continuation determined at specific points during its lifetime \cite{carson_421_2013}.


    In conjunction with the waterfall model, incremental or iterative practices are often used \autocite{HEEAGER201822}. \textbf{Incremental} development refers to a method where distinct parts of the design are delivered one at a time; \textbf{iterative} development refers to an approach where the complete system is being refined after each of many cycles --- the goal being that each cycle leads to an improved product, by using the outputs and lessons-learned from the previous cycles \autocite{HEEAGER201822}. More systems engineering concepts have been extensively explored, such as the spiral model \autocite{nasa_engineering__safety_center_aligning_2018}, concurrent engineering \autocite{bandecchi_concurrent_1999}, and Model-Based Systems Engineering (MBSE) \autocite{fischer_implementing_2017}. It is also common to use a \textit{hybrid} approach, combining together different methods \autocite{carpenter_is_2014,garzaniti_toward_2020}, for example by using different product management techniques in different phases/parts of the project.

    These traditional approaches have seen success with large, monolithic missions \autocite{shea_nasa_2017, HEEAGER201822}, especially with safety-critical systems \autocite{kasauli_safety-critical_2018}. However, smaller spacecraft like CubeSats have characteristics (such as lower complexity, lower costs, shorter schedules, higher risk acceptance, easier integration, smaller teams) which do not always justify the overhead added in terms of cost, time and resources added by naively implementing traditional methods \autocite{carson_421_2013,yassine_information_2003,royce_managing_1970,sebok}. 

    \paragraph{}

    In CubeSats, different approaches can be followed, depending on the nature of the project (commercial or educational) and the structure of the team:
    \begin{itemize}
        \item \textbf{Sequential approaches}: The typical progression of phases is followed (concept, design, assembly, verification and operations), but with modifications or shortening of each project phase. Usually phases 0, A and B (from conception until preliminary design) are combined into a single Phase AB. In some cases, a demonstration of the mission feasibility using a prototype may be required from Phase AB \autocite{nieto-peroy_cubesat_2019,lubian-arenillas_nanosatellite_2019,tyvak_trestles_2021}.
        \item \textbf{Evolutionary approaches}: Many CubeSat projects follow iterative or incremental approaches, where design, development, and testing may happen concurrently or repeatedly \autocite{lubian-arenillas_nanosatellite_2019,alanazi_engineering_2019,sousa_cubesat_2021}. For example, it is common to manufacture engineering models or other representations of subsystems, before proceeding to system-wide assembly \autocite{faure_toward_2017}. It is also common to work in an iterative approach for the entire system, by producing different ``versions'' of a CubeSat \autocite{decker_systems-engineering_2016,cappaert_building_2018}. Iterative development can also happen by working on a reduced version of the entire satellite before assembling; this is often implemented through the ``FlatSat'' approach, where subsystems can be tested long before feature completeness \autocite{nasa_cubesat_launch_initiative_cubesat_2017,MAIVP}. Software, more specifically, can be tested continuously and independently from the rest of the subsystems, leading to quick verification \autocite{kiesbye_hardware---loop_2019}.
        
        \item \textbf{Emergent approaches}: CubeSat projects have applied approaches fitting the term ``agile'' \autocite{angeli_paving_2014,labarge_cubesat_2014,lubian-arenillas_nanosatellite_2019}, where the focus is shifted from requirements compliance to human interactions and customer satisfaction \autocite{beck_agile_2001}. More approaches again focus on agile software development \autocite{lill_agile_2018,coyle_eecsat_2020}.

    \end{itemize}

    \label{sec:approaches}

    \paragraph{}

    Pico-satellites, and especially femto- and atto-satellites, have a number of characteristics and capabilities (notably low cost, small size, fast development, mass production, potential for rapid technology testing, small development teams \autocite{hein_attosats_2019}) that call for new systems engineering methods, specific to them.
    
    It is generally accepted that plan-driven approaches are better suited for large projects and teams, which require low risk and predictability \autocite[Sec. 2]{boehm_balancing_2004}. In contrast, sub-CubeSat spacecraft are small developments that can benefit from quickly responding to change, reduction of risk through redundancy, and small, knowledgeable teams. In this case, a traditional approach would add costly overhead in information sharing \autocite{yassine_information_2003}, unnecessary bureaucracy \autocite{boehm_balancing_2004}, high inertia \autocite{carson_421_2013}, and would not adapt quickly to new technological developments \autocite{sebok}.

    \paragraph{}

    In this work, we claim that:
    \begin{enumerate}[label={\alph*)},itemsep=0pt]
        \item By creating a tailored methodology, we can produce femto- and atto-satellites in only a few months, with a development cost at least an order of magnitude lower \todo{how to give evidence?} than one of a CubeSat, and with minimal schedule overruns.
        \item Already existing systems engineering approaches for simple systems can be easily tailored to a femto- or attosat, in order to satisfy the previous claim.
    \end{enumerate}  


\section{Research Method}

To resolve the claims of the previous section, we will propose a systems engineering method \autocite{gerrike_what_2017} tailored to femtosatellites and attosatellites.

In order to gain data and verify our method, we performed a Descriptive Study \autocite{blessing_drm_2009}. The nature of space systems development means that it is difficult to repeat similar conditions in a controlled environment. At the same time, due to the novelty of sub-CubeSat spacecraft, there is not enough available information to perform statistical analyses. Therefore, we applied the \textit{case study} data collection method \autocite{blessing_drm_2009,yin_case_2014}, by following two case studies of miniaturised missions in UL (\Cref{sec:caseintro}).

In order to develop our method, we will start by investigating popular frameworks for rapid prototyping and development, which can be easily tailored to the characteristics of sub-CubeSat spacecraft. After selecting a framework that seems best suited to these characteristics, we will apply the lessons learned from the two case studies, as well as suggestions from the literature, in order to create a more well-tailored method (\Cref{sec:agilee}).

\subsection{Case Studies}
\label{sec:caseintro}

The two case studies selected are missions that, as of 2022, are under manufacturing in the University of Luxembourg (UL) by the SpaSys team. The missions are two femto platforms that are used for in-orbit technology demonstration:
\begin{enumerate*}[label=\alph*)]
\item A 10 \(\times\) 10 cm payload, used to test Artificial Intelligence for thermography, and
\item A 5 \(\times\) 5 cm chipsat, used to test visible light communication.
\end{enumerate*}

While the first mission is a payload and not an independent satellite per se, the characteristics of its development are similar to the ones of an attosatellite, as there is limited functionality that needs to be achieved with low cost, but there are significant constraints in terms of available space, data budget and interfaces.

 Our selected case studies are differing in type, technologies used, complexity, fractionation and team structure. This diversity is useful to evaluate if our method can be generalised to different spacecraft.
We can then test the \textit{external validity} \autocite{yin_case_2014} of these case studies, i.e.~we will use theoretical results to generalise from their specific findings to generic femto- and atto-satellites.

\subsubsection{AI4Space}

\begin{figure}[h]
    \centering
    \includegraphics[width=.8\linewidth]{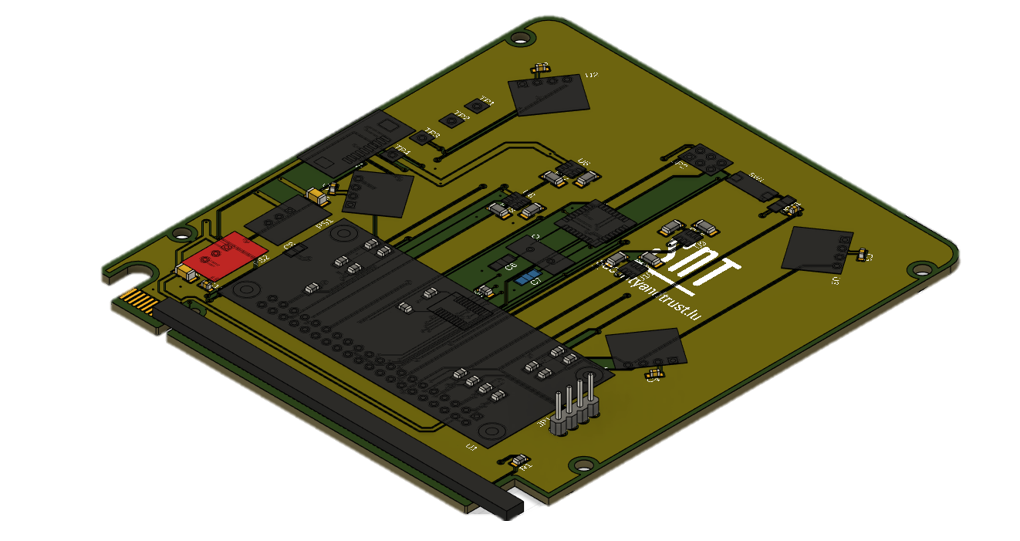}
    \caption{Render of the AI4Space payload}
    \label{fig:ai4space}
\end{figure}

The objective of the AI4space mission is to launch computer vision algorithms in space and serve as a testbed for the end-to-end development of space systems in UL, in collaboration with the CVI2 research group. 
The compact-infrared payload of the AI4Space mission aims to detect thermal anomalies on space electronics hardware. The thermal anomalies are detected using Infrared Thermography and Artificial Intelligence. Infrared thermography is a non-invasive method that uses infrared cameras to detect temperature variations of electronic components.

\todo{Please check if the attributions (spasys, UL, CVI2) are correct.}


The payload is a PCB (Printed Circuit Board) and has a size of 100 mm \(\times\) 100 mm \(\times\) 15 mm and it is hosted by the Skyride payload hosting programme of \href{https://www.skykraft.com.au/}{Skykraft}. It has the following subsystems:

\begin{itemize}[itemsep=0pt]
    \item A Raspberry Pi, containing the onboard software developed in Python.
    \item A ``mothercraft'' interface, exchanging telecommands and telemetry with the Raspberry Pi.
    \item A power interface, supplying power to the payload from the mothercraft.
    \item An Arduino, acting as a target board that shows temperature variation when its clock frequency is modified.
    \item Heaters, used to increase the temperature significantly to be captured by the infrared cameras.
\end{itemize}

\subsubsection{ChipSat}

\begin{figure}[h]
    \centering
    \includegraphics[width=.6\linewidth]{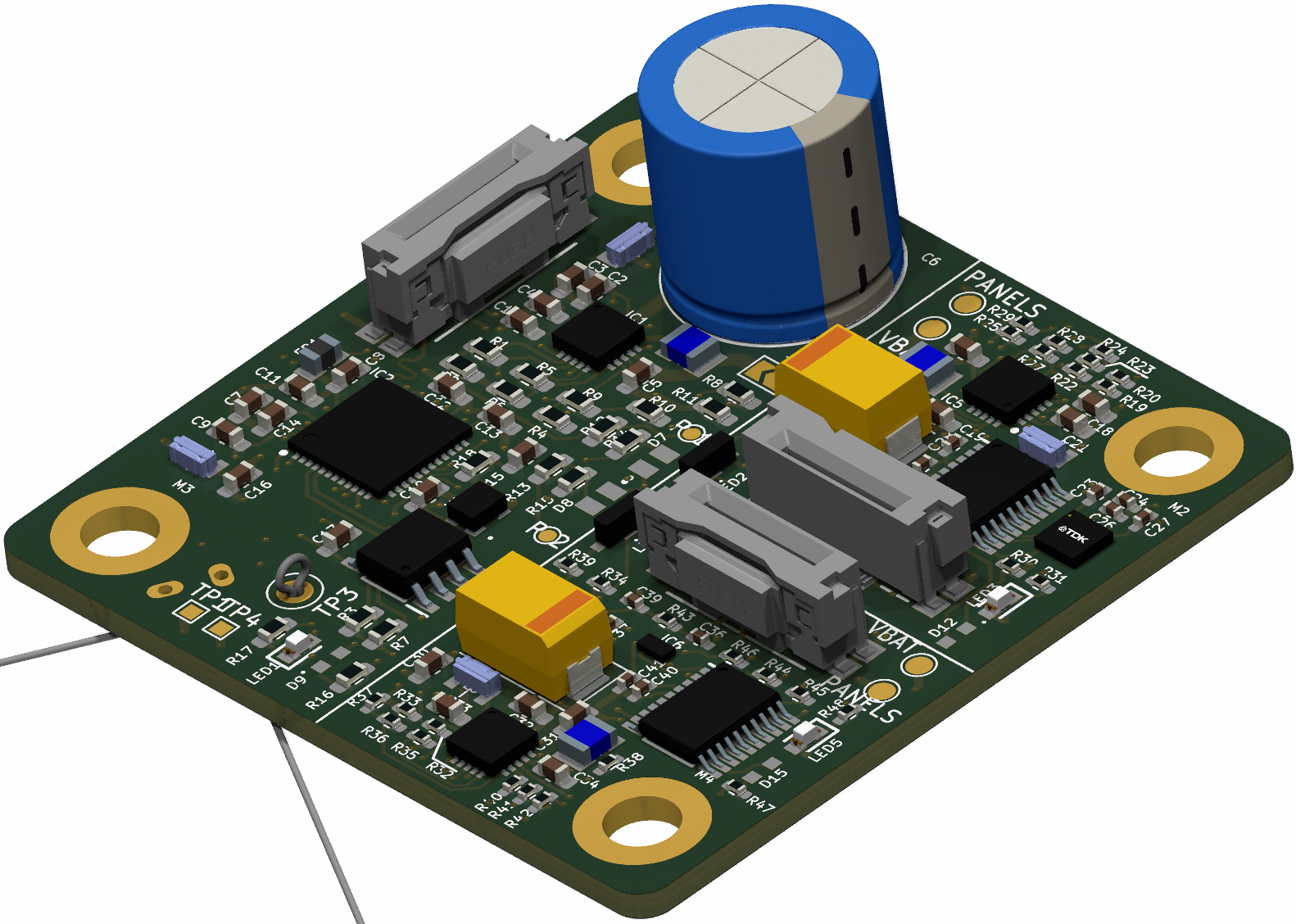}
    \caption{Render of the ChipSat PCB}
    \label{fig:chipsat}
\end{figure}

 The ``ChipSat'' mission consists of a \acs{PCB} containing 3 independent attosatellite designs, which will be mounted on the side of a microsatellite. The objective of this mission is to investigate the feasibility of free-space visible optical light communication between different satellites of a fractionated system. The ``network'' consists of two types of satellites: a ``primary'', responsible for communication with the ground, and a ``secondary'', responsible only for data generation. For its maiden flight, the ChipSat will be mounted on a microsatellite, but electrically independent from it.  More specifically, the contained components are:
 \begin{itemize}[itemsep=0pt]
     \item Microcontrollers, containing the onboard software developed in C++. For the primary, it is an \acs{SoC} that contains an \acs{RF} transceiver.
     \item LEDs and photodiodes, used to perform low-rate visible communication.
     \item Miniaturised solar panels, connected to an energy harvester \acs{IC}, generating an average sunlight power of \textasciitilde\SI{200}{\milli\watt}.
     \item For the primary, a deployable dipole antenna for low data-rate signals.
     \item For the secondaries, a gyroscope and an ambient light sensor.
 \end{itemize}

\subsection{Agile}
\label{sec:agilee}

    Out of the methodologies analysed in \Cref{sec:approaches}, we focused on ``emergent'' approaches, which best suit ChipSat characteristics \autocite{sebok}. These approaches are usually grouped under the term ``agile'', which originates from the \emph{Manifesto for Agile Software Development} \autocite{beck_agile_2001}, published in 2001.


    The Agile manifesto aims to increase productivity by pursuing the following four values: \autocite{beck_agile_2001}
    \begin{enumerate}[itemsep=0pt]
        \item \emph{Individuals and interactions over processes and tools}
        \item \emph{Working software over comprehensive documentation}
        \item \emph{Customer collaboration over contract negotiation}
        \item \emph{Responding to change over following a plan}
    \end{enumerate}
    Agile approaches therefore often follow an iterative cycle \autocite{HEEAGER201822}, each step of which results in a deliverable, working product. 

    While Agile originates from Software Engineering and has been observed to improve the effectiveness of engineers \autocite{noauthor_15th_nodate,noauthor_status_nodate}, it was quickly generalised to systems engineering \autocite{haberfellner_agile_2005}, citing improved engineering efficiency, early Return On Investment, responsiveness to change and increased project control \autocite{douglass_agile_2015,kohlbacher_agile_2011,darrin_agile_2017}.

    \colorlet{No}{MaterialRed600}
    \colorlet{Yes}{MaterialGreen600}
    \def\yes{\textcolor{Yes}{Yes}}
    \def\no{\textcolor{No}{No}}
    \begin{table*}[h]
        \caption{Comparison of popular agile frameworks \autocite{boehm_balancing_2004,dingsoyr_decade_2012}}
        \label{tab:compagile}
        \label{sec:scrum_intro}
        \def\arraystretch{1.4}
        \begin{tabularx}{\linewidth}{@{}L{5cm}@{\hskip 1em}L{1.3cm}L{1.3cm}L{1.3cm}@{\hskip 1em}X@{}}
        \toprule
         & Adaptable to hardware & Full lifecyle & Small systems & Comments \\ \midrule
        \textbf{Kanban} & \yes & \no & \yes & Does not explicitly address verification/testing \\
        \textbf{Scrum} & \yes & \yes & \yes &  \\
        \textbf{Lean   Development (LD)} & \yes & \no & \yes & Strategic, risk-driven approach, not focused on systems engineering \\
        \textbf{Crystal   Clear} & \yes & \yes & \yes & "Crystal" method for very small teams \\
        \textbf{eXtreme   Programming (XP)}&\no&\no&\yes &  \\
        \textbf{Dynamic   Systems Development Method (DSDM)} & \yes & \yes & \no & Closer to plan-based ``traditional'' methods, emphasis in management activities \\
        \textbf{Feature-Driven   Development (FDD)} & \no & \no & \no & Mostly focused on individual practices for software development \\ \bottomrule
        \end{tabularx}
    \end{table*}
    
    Practices considered to be agile have already been considered and used, partially or completely, in various space missions \autocite{carpenter_is_2014, dillon_faster-better-cheaper_2015, carson_421_2013} and especially in CubeSats \autocite{honore-livermore_agile_2021,dallmann_agile_2015,lill_agile_2018,berthoud_university_2019}. Agile is often used in small teams with limited resources which need turnkey developments and do not have significant risks involved \autocite{carson_421_2013}.


    \paragraph{}

    Agile practices are usually implemented through specific frameworks. In this article, out of the rigorous frameworks defined in \textcite{boehm_balancing_2004}, we are focusing on the most popular ones \autocite{noauthor_status_nodate}, shown in \Cref{tab:compagile}.
        



    In order to choose a baseline to develop a tailored method, we selected a network out of those based on the following criteria:
    \begin{itemize}
        \item \textbf{Easy adaptability to hardware}: Most Agile frameworks are specifically built around software-based practices and tools, such as daily unit testing and instant integration, which cannot be implemented in a larger system. In this article, we consider frameworks that can be easily applied to physical systems.
        \item \textbf{Covering full lifecycle}: Many Agile frameworks cover only certain parts of a product's lifecycle, or provide solutions only for one specific project management aspect. For convenience, we consider frameworks that cover aspects from planning to verification.
        \item \textbf{Small systems}: Some Agile frameworks are built for larger systems with higher complexity and larger teams. In this work, we consider frameworks that claim to work for \textasciitilde 10 maximum people.
    \end{itemize}

    From \Cref{tab:compagile}, it seems that \emph{Scrum} and \emph{Crystal Clear} are the most suitable frameworks for our proposal. In this work we will focus on \emph{Scrum}, because of its very high popularity and available literature \autocite{noauthor_status_nodate}. Scrum has also specifically been proposed for use in Systems Engineering, especially for systems that are accepting to rapid changes, by \citeauthor{bott_analysis_2019} \autocite{bott_analysis_2019}.
    
    However, we do note that our search is by no means exhaustive; other frameworks may also present relevant opportunities for tailoring.





\section{Applying Agile to sub-CubeSat spacecraft}

In order to develop the new method, we will apply the characteristics of sub-CubeSat spacecraft to published findings related to agile systems engineering. We will also use the lessons learned from the two case studies (\Cref{sec:caseintro}).

\subsection{Distinctive femtosatellite characteristics}

To prepare a set of guidelines for the use of Agile in sub-CubeSat developments, we first present some of the most notable distinctive characteristics of such spacecraft \autocite{hein_attosats_2019,barnhart_low-cost_2009,manchester_centimeter-scale_2015}:
\begin{itemize}[itemsep=0ex]
    \item Low design complexity 
    \item Low manufacturing cost
    \item Low launch cost, due to low mass
    \item Femtosatellites and smaller spacecraft often have little distinction between subsystems in the traditional sense. Using \acf*{SoC} technology, multiple or even all subsystems can be combined into a single component \autocite{wolf_multiprocessor_2008}.
    \item Technology reuse is not yet explored in detail. In contrast to the CubeSat standard and the relevant deployer specifications and component market \autocite{CDS14}, femtosatellites and smaller are not supported by off-the-shelf specifications or modules as of 2022. However, standardisation for PocketQubes is already underway \autocite{PQ1, PQ91}.
    \item Interface requirements for the deployer and/or launcher are invariants and will usually serve as the main drivers for the design.
    \item Engineering teams working on sub-CubeSat spacecraft usually consist of few people, and are not spread into different subteams.
    \item The facilities for manufacturing and testing such spacecraft can often be available in-house \autocite{triantafyllopoulou_qubik_2020}.
\end{itemize}

\subsection{Agile principles} \label{sec:agile_principles}

\citeauthor{douglass_agile_2015} (\citeyear{douglass_agile_2015}) \autocite{douglass_agile_2015} provides seven core ideas for agile methods. In this section, we will restate and analyse their applicability to sub-CubeSat projects.

\subsubsection{Work Incrementally}

If a product is separable into distinct parts, which can be developed, tested (and preferably deployed) independently, then the work can be easily divided into multiple cycles.

When a system is considered ``complete'', i.e.~it is operational and fulfils the basic objectives of a mission, incremental development can turn into iterative development, where the focus is on the rework and improvement of the existing system, rather than the implementation of critical missing features \autocite{HEEAGER201822}. This is especially applicable to femto- and attosats, as the time between conception and manufacturing can be very short.

The time between each cycle varies between each projects. While 2 weeks are common in software engineering projects, systems where hardware and physical effort is involved may benefit from longer iterations (4--6 weeks).

\subsubsection{Plan Dynamically}

A common myth is that Agile practices call for no planning or bureaucracy \autocite{douglass_agile_2015}. \textcite{douglass_agile_2015} claims that \textit{``planning is important but only if it is accurate''}. Agile principles introduce a degree of uncertainty in all layers of a design, from customer requirements to implementation. The project plan therefore should be defined, but also dynamic: it should reflect this uncertainty and be frequently updated (e.g.~every 1--2 months) to improve the assumptions made about the needed amount of work and ``velocity'' of the team.

\subsubsection{Actively Reduce Project Risk}

In terms of project planning, uncertainties can introduce risk. Design changes or active actions can help to mitigate risks that are identified early or late in the design.

In our experience, focusing on the critical path in terms of schedule, or the components with the highest impact in the system, will help identifying the most threatening risks. For example, any interaction with an external supplier or provider is coupled with uncertainty and delays, and should hence be started as early as possible during the project. To mitigate against a possible chip shortage, designers should procure critical and irreplaceable components as early as possible (even during the design), and maintain a reliable stock.


\subsubsection{Verify Constantly, Integrate Continuously, Validate Frequently} \label{sec:cicd}

Verification in space systems is usually formal and follows a unit--subsystem--system path \autocite{bundesrepublik_deutschland_v-modell_2006}. On the other hand, in software systems, verification is done through automated test suites, and can be completed in just a few minutes \autocite{beck_test_2002}.

While it is easy to test spacecraft software, especially when complemented by a hardware-in-the-loop environment \autocite{fritz_hardware---loop_2015}, the engineering effort to manufacture and test space hardware can often take years. This reduces the responsiveness to change and the chance to identify errors early in the design \autocite{garzaniti_effectiveness_2019,peterson_when_2021}.

However, when developing picosatellites and smaller spacecraft, the manufacturing and assembly can be made in a few months with little overhead. This means that \textbf{the verification of representative hardware can become an important part of the design process, as verification can happen at the end of each development cycle}. The hardware used in these cases could usually resemble a Structural Model, a Development Model or an Engineering Model \autocite{ECSS-E-HB-10-02A}. Engineering Models in particular should be representative in terms of form, fit and function, without requiring hardware and processes suitable for space.

Especially for femto- and attosattelites, low launch costs mean that there is minimal effort required to reach orbit. \textbf{A system can be iteratively verified by sending its different versions into orbit.} Lessons learned from the actual operation of a satellite will then be used to improve the next iterations. In software engineering, this concept is often referred to as \textbf{Continuous Integration/Continuous Delivery (CI/CD)}, and involves constant releases of a product which, while not perfect, are fully functional and operational \autocite{beck_extreme_2008,shahin_continuous_2017}.

Alternatively, as a middle-ground solution, high-altitude balloon flights, or even actual usage of the spacecraft on the ground, could be implemented \autocite{adams_theory_2020}.

Given the above, we can summarise the different verification methods for sub-CubeSat spacecraft: \autocite{ECSS-E-HB-10-02A}
\begin{enumerate}[itemsep=0pt]
    \item Analysis and Review of Design
    \item Testing (software only)
    \item Testing (software and hardware)
    \item Field operations (on-earth demonstrations, suborbital flights)
    \item In-flight operations
\end{enumerate}

We note that, based on the available resources and objectives, a different verification method out of the above can be chosen for different cycles, phases or parts of the same project. We also note that this sequence does not need to be followed in order; for example, after a test flight, the team may continue with hardware testing of the next iterations.

Validation \autocite{ECSS-E-HB-10-02A}, usually in collaboration with the customer, also happens regularly in the context of Agile.






\subsubsection{Modeling Is Essential for agile MBSE}

\textbf{Model-Based Systems Engineering (MBSE)} practices can be applied to agile systems engineering, especially if they do not need considerable overhead and are easy to implement \autocite{tang_mbse_2018,douglass_agile_2015}.

For example, an interesting concept in software engineering is that of \textbf{Test Driven Development (TDD)}. In TDD, automated tests are written before the development of the code itself \autocite{beck_test_2002}. The task of the developer is then to make sure that the developed system passes all the tests. Consequently, in TDD, the test itself serves as a specification or a model for the requirements of a system.

Models can also be used for automated generation of software \autocite{perrotin_taste_2012,bychkov_using_2018}, or for the reduction of the work needed to document a system. However, mechanicals and hardware \autocite{de_vos_documentation_2022} still require manual effort to integrate into a model.


\begin{figure}
    \centering
    \includegraphics[width=\linewidth]{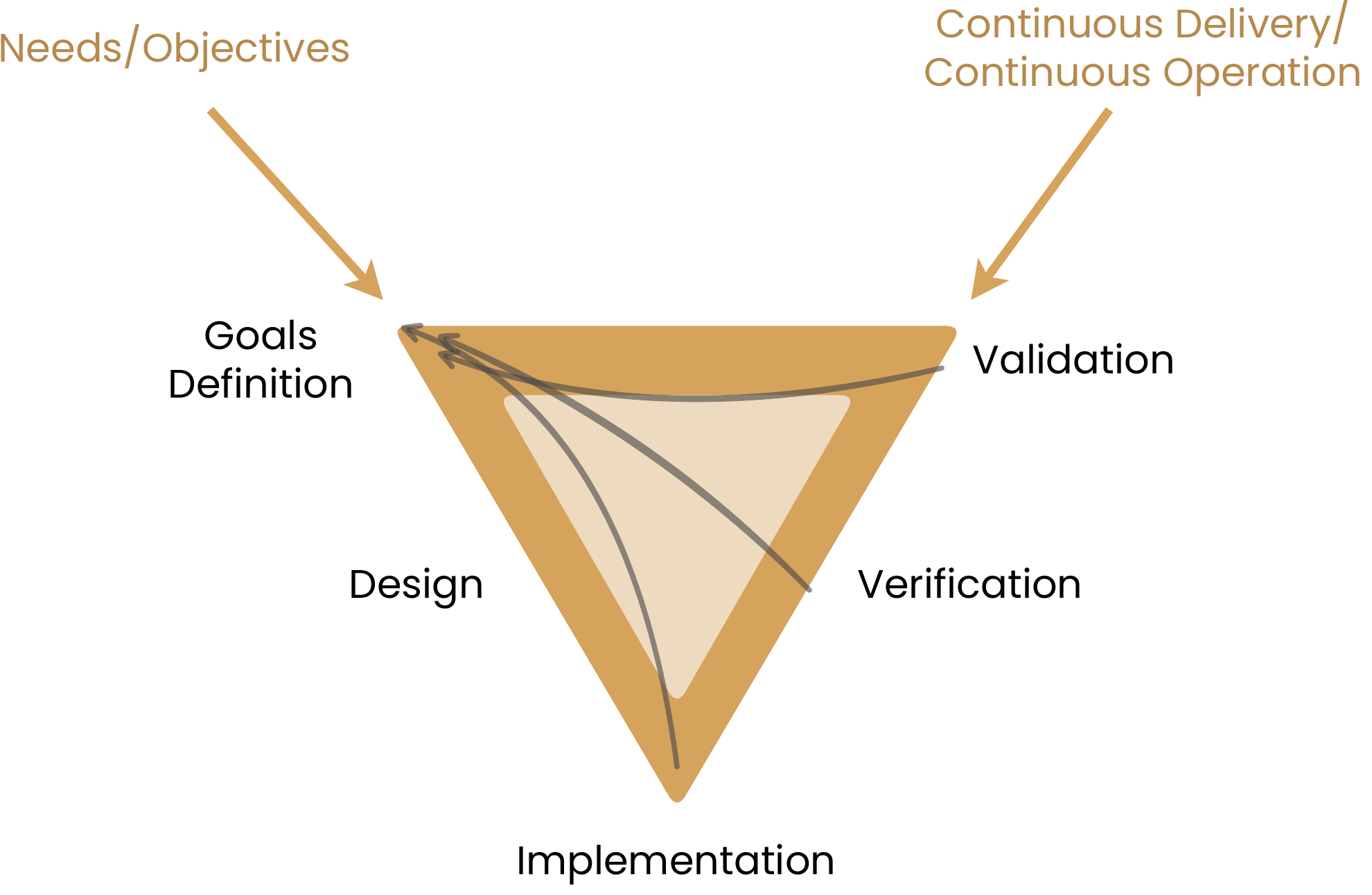}
    \caption{The triangle model}
    \label{fig:nabla-model}
\end{figure}

\section{Proposed Methods}
\label{sec:proposed_methods}

In this section, we will propose two agile methods that follow from the analysis performed in the previous paragraphs. The first one is the result of getting inspiration from the traditional V-model, and modifying it with some additional flexibility, so that it meets the definition of agile. The second one starts by directly implementing the Scrum agile framework, and applying to it the constraints of a sub-CubeSat spacecraft.

\subsection{The Triangle model}
\label{sec:the_triangle_model}

In the core ideas described in the \hyperref[sec:agile_principles]{previous section}, the overarching theme of \textbf{iterative development} is dominant. 
We can therefore adapt the V model into a \textbf{triangle model}, where paths exist for \textbf{going back to the conceptual stage} at every point during the design. We will use this triangle model as a representation \autocite{gerrike_what_2017} for the proposed model.

Instead of requirements, the triangle model originates from \textbf{needs and objectives}. We define these as qualitative targets that a customer or entity has for a mission, such as \emph{creating space heritage within an organisation}, or \emph{measuring temperature variations during reentry}.

Given the needs and objectives, the more specific and quantitative goals are stated later. We still refrain from using the word ``requirements'', as we want to avoid the rigour and effort connected with defining low-level requirements at the beginning of a project. However, we note that these goals still need to be well-defined, and can refer to the mission, the system, or specific units. For example, a goal would be to \emph{take at least 10 measurements during reentry}, \emph{develop a spacecraft with a \(<\)25\si{\gram} mass}, or \emph{have a data rate of at least 100\si{kb/s}}.

\paragraph{}

The next steps in the process follow the design and implementation. We deliberately did not distinguish between \emph{high-level} and \emph{low-level} design, or \emph{subsystem-level} and \emph{component-level} design. This is done because pico-scale spacecraft can incorporate different subsystems into single components or units. It is up to each team to determine if they will split the design into multiple levels, based on their size and objectives.

Verification and validation follow after implementation. Validation marks the end of one iterative or incremental cycle. The developers can then proceed with the operation of the validated product (\Cref{sec:cicd}), and they can start the next cycle from a refinement of the initial goals.

\paragraph{}
However, the triangle model does not only follow this circular path: Any point in the triangle can lead to \textbf{back-tracking}, i.e. going back to the goals definition. This is essentially the \textbf{mechanism to respond to changes} during the design. For example, a team might start with a goal of \emph{reaching a data rate of at least \SI{100}{\kilo\bit/\second} with a \SI{5}{}\(\times\)\SI{5}{\centi\meter} attosat}. During the implementation phase, they might find out that their radio transceiver is too large to fit on the attosat's PCB.
They will then be forced to backtrack, and change their goals to increase the satellite's size, or support slower data rates with a smaller transceiver.

A significant difference between typical incremental models \autocite{sebok} and the triangle model is that the latter does not require going through all the phases before back-tracking: Lessons learned during design or implementation may immediately lead to a redefinition of the goals, without a need to go through the next phases.

An implementation of the triangle model also does not need to begin from the top left. The rapidly-developing COTS component ecosystem means that, for example, a proposed implementation may be available before a specific use case has been prepared. In this case, the system becomes largely \textbf{capability-driven} instead of goal-driven.

\subsection{Sat-Scrum}
\label{sec:sat_scrum}

\begin{figure*}
    \centering
    \includegraphics[width=\linewidth]{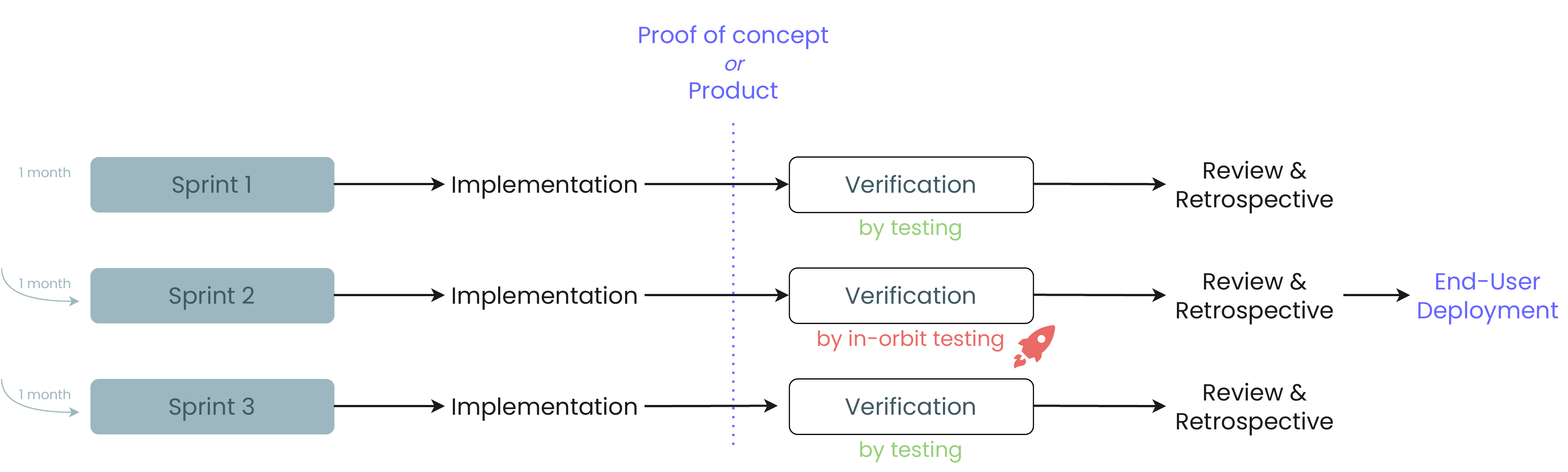}
    \caption{Example of applying Scrum to sub-CubeSat scale space projects}
    \label{fig:scrum}
\end{figure*}

While the Triangle Model proposed in \Cref{sec:the_triangle_model} presents significant differences in terms of flexibility in project management, it is still covering a high level of the design process, and, as a tailoring of a more ``traditional'' process, may not fully take into advantage the capabilities offered by miniaturised spacecraft.

Therefore, continuing the analysis presented in \Cref{sec:agilee}, in this part we will focus on \emph{Scrum}, and how it can be adapted to our target platform.

Scrum is a lightweight agile framework that covers all life cycle activities of a project or a system, and is based on empiricism, focusing on transparency, inspection and adaptation \autocite{schwaver_definitive_2020,bott_analysis_2019}.


The definitions of Scrum \autocite{schwaver_definitive_2020} contain a lot of nuance and terminology that is suggested to be followed by participating projects, to encourage a shift in mentality from a traditional plan-based approach. Here we will focus on the idea of \textbf{Sprints}, which correspond to the cycles of an incremental/iterative process. A sprint starts from a backlog (prioritised list of work items) and results in a sprint goal, which might be the next version of a product.
A sprint usually includes:
\begin{itemize}[itemsep=0pt]
    \item \textbf{Sprint Planning}, a long meeting where a goal and plan is laid out for the next sprint
    \item \textbf{Daily Scrums}, which are 15-minute (timeboxed) meetings that take place to gauge progress and update the short-term planning.
    \item \textbf{Sprint Review}, a technical session at the end of a sprint, where its outcome is reviewed and technical lessons learned are assessed.
    \item \textbf{Sprint Retrospective}, a meeting called after the Review, where the focus is on interactions, processes and tools, and how the team can solve problems that occured in the future.
\end{itemize}

Scrum also defines the positions of the ``Scrum Master'', the Product Owner and the Developers, the last two of which are the main responsibles for selecting the backlog work items.

\paragraph{}

Applying the idea of Sprints to a space project, in the same vein as the triangle model, can result to something similar to what is shown in \autoref{fig:scrum}.

In this case, every sprint results in a functional, improved product.
During development, verification can be done through one of the five methods listed in \Cref{sec:cicd}. In this case, even \textbf{in-orbit testing can be part of the normal product development procedure} in each sprint. Flight models do not need to be ``feature-complete'', only operational; this is aided by the low resources that mainly femto- and attosats need to be launched, essentially if the manufacturer takes advantage of economy-of-scale effects \autocite{hein_attosats_2019}.

Due to the complexity of the systems, the duration of a sprint in this case will lean towards the longer end of \SIrange{4}{6}{weeks}. The verification process also needs separate considerations, since it might not be able to be fully automated as in a software system.

In the case of physical hardware, verification might need some days or weeks to be performed, as an activity separate to development. It might include time spent for procurement or assembly. In specific project stages or situations (e.g.~breadboard models, or modular assemblies) these activities might be able to be completed in less than a week, in which case they can be a regular part of the sprint.

However, larger developments (e.g.~producing Engineering Models) or tests (e.g.~test flights) might require significantly more involvement and cannot be completed as part of a single sprint. In this case, the ``review \& retrospective'' parts for every sprint cannot fully reflect the verification done in the sprint. However, they lessons learned and results from this process can be implemented in future sprints.

\paragraph{}
A few other qualities of Scrum that might be interesting to analyse are:
\begin{itemize}
    \item In Scrum, there is no clear separation between different functions of the developers: Design, development, quality assurance or analysis. Team members are self-managing and distribute work items based on their discretion.
    \item Scrum encourages transparency and information sharing for all parts of the work. This can be achieved by using the appropriate project management tools, having a central archive to gather results and information, avoiding personal communications, or even by sharing a common workplace.
    \item We note the differences between `incremental' and `iterative' development, as defined by \citeauthor{HEEAGER201822} \autocite{HEEAGER201822}. In the beginning of the project, a ``scrum increment'' will be identical to an ``agile increment'', as parts of the system will be built for each Sprint until it is functional. However, in the largest part of a project, a ``scrum increment'' will more closely match an ``agile \emph{iteration}'', where the focus is on the rework of the system so that it implements more of the defined goals.
    \item Especially in a space system, the complexity of managing different increments, versions, physical products, test campaigns and even launches, would require significant \textbf{tooling} to remove the management overhead and make access to information easy. A combination or adaptation of tools already available in the software engineering, project management and space systems engineering might be useful for this purpose. For example, a Version Control system could be used to store different versions of software, designs or documentation; a Project Management tool built for Scrum can support the work allocation and planning for team members \autocite{ciancarini_open_2020}; and Model-Based Systems Engineering can be used to formalise the most crucial and high-level parts of the process \autocite{boggero_mbse_2021,douglass_agile_2015}. However, it is important that the tooling used does not act as an impediment to the daily work of the developers.
\end{itemize}



\section{Case Studies}

In this section, we will provide details on the Case Studies presented in \Cref{sec:caseintro}. We will start by presenting the development flow of these missions, as well as some practices that were followed and lessons learned. We will finally emulate the application of the methods of \Cref{sec:proposed_methods}, following a ``counter-factual'' approach, since both missions already started before these methods were developed \autocite{hein_evaluating_2020}.
\todo{Too informal?}



\subsection{AI4Space}
\label{sec:ai4spaceresults}


The project began with understanding the needs and objectives of the AI-driven payload. Then, the software and hardware architecture were documented to satisfy the space mission objectives. While we initially used MBSE to model requirements during the development phase, there was not enough time to complete and utilise this formalisation. COTS components were chosen instead of space-grade products since they were easily available, less expensive and reduced software and hardware development time. Additionally, ready-to-use development platforms such as \href{https://www.edgeimpulse.com/}{Edge Impulse} were used to embed AI algorithms on the computer module to reduce development time. Finally, the functionalities of the payload were tested using a breadboard model to validate the proof of concept, before building the qualification and flight models. 





For the development of AI4Space, we experimented with a tailored approach that incorporated various iterative practices adhering to the agile manifesto \autocite{beck_agile_2001}, but without belonging in a concrete framework:
\begin{itemize}
\item We designed a system that was \textbf{shippable from the early stages}. We implemented the basic operational functionality with high priority in the first few months. This included basic operational functions, such as telemetry and telecommands, downlinking science data, or command scheduling (``time-tagging''). It also included at least a proof-of-concept for the payload, with the basic operations and sensor code being written only over the first 2 weeks after development started.
\item \textbf{Multiple functional iterations} were made during development.
We roughly followed the ``Sprint'' pattern, performing one iteration every 2 weeks, which was concluded with a review and planning meeting at the end of the sprint.
\end{itemize}

In order to manage all of the above, a hybrid method was used:
\begin{itemize}
\item The major mission milestones were managed in a traditional way, following a \textbf{Gantt chart}, and splitting the mission into specific large work packages, starting from \emph{Preparation} until \emph{Scientific Exploitation} of the results.

\item For the design and development, \href{https://gitlab.com/}{\textbf{GitLab}} as a tool for project management, work item tracking, and result sharing.
\item For the assembly, we created a standard assembly procedure which included the list of the required components. Hardware stock tracking was done using \href{https://inventree.readthedocs.io/en/latest/}{Inventree} to monitor the components purchased, current stock and components to order. 
\end{itemize}

While we could easily adopt agile practices for the payload's software, the constraints imposed by hardware (mainly cost, time and part availability) are harder to work with \autocite{garzaniti_effectiveness_2019,peterson_when_2021}. In the project's context, we worked simultaneously on a functional \textbf{Breadboard Model} that was developed incrementally, and the PCB designs for the \textbf{Qualification/Flight} models. The following principles were followed throughout development:
\begin{itemize}
    \item ``Stand-in'' parts were used when a component or hardware function was not readily available. We deemed it more important to have a \emph{functional} instead of a \emph{representative} model through the early stages. For example, we used a Raspberry Pi 3 instead of a Raspberry Pi 0 for the early days of development.

    It is important to note that the loss of representativeness has to be recorded and known in the team, so that fewer surprises show up later in the process. In some cases, it might even be possible to emulate the differences in the configuration, for example by limiting the amount of memory available to the Raspberry Pi in software.
    \item The software was implemented in conjunction with the hardware. 

    Specifically for AI4Space, our main platform was a Linux distribution on the Raspberry Pi. While it was possible to develop most parts of the software independently from the hardware ("off-line"), we opted to validate the written software on the hardware models immediately, as part of a CI/CD process.
    \item \acs*{MBSE} was used for the definition of interfaces. The ``model'' in this case was Python code, which included the formal definition of the interfaces, along with documentation. Python's \texttt{construct} library \autocite{bulski_construct_2020} could then immediately parse and generate telemetry \& telecommands, using only this single source of data. We also prepared a script that automatically created human-readable documentation for this interface.
    \item Parts of the system were tested automatically, using the Robot automation framework \autocite{robot_framework_foundation_robot_nodate}. The specific framework is not software-centred and provides a user-friendly test format, inspired by the concept of ``user stories'' in Behavior-Driven Development \autocite{north_whats_2007}. This means that:
    \begin{itemize}[itemsep=0pt]
        \item Testing is not limited to software functions only; An appropriately configured system can also perform hardware verification.
        \item Testers can be independent from the developers, since our tests were reasonably decoupled from the code they test.
        \item Test stories could also be used for nominal procedures, such as clearing spacecraft logs.
        \item There can be an additional slight learning curve, since developers need to become familiar with the domain-specific language introduced by the framework.
    \end{itemize}
\end{itemize}

During development, we also observed that:
\todo{Should I just combine the lessons learned from each mission into a single section?}
\begin{itemize}[itemsep=0pt]
    \item The nature of the team meant that members had to spend a significant amount of their time in tasks unrelated to the project.
    \item Not enough time was dedicated to documentation, which steepened the learning curve for new members. However, the fast pace of development meant that any written documentation would very quickly become obsolete, even if it was written at a high level only. Ideally, the appropriate level of documentation would be written as a required part of development, or the use of automated documentation generation tools could be extended.

    Documentation need not only explain the function of a system; it can also contain information about the design justification. In our cases, this was done on a rolling bases: When opening and closing an issue on Gitlab, we added at least a short sentence explaining the reason for this action. 
    \item Some unplanned events caused ``road-blocks'' that had to halt development and break momentum, such as procurement delays, supplier stock issues, and equipment or network issues. After these, we devised a ``Minimum Working Environment'' that required only a personal computer for a developer to work towards their goal.
    \item For software development, we followed an empirical approach instead of laying out a detailed plan beforehand, and we did not over-design, meaning we did not add more functionality, modularity or levels of abstraction, apart from what was needed in each iteration. This meant that parts of the code had to be \textbf{refactored} in some situations. 

    While detailed planning might prevent time spent on refactoring, the Scrum Guide \autocite{schwaver_definitive_2020} suggests that empricicism and lean thinking reduce wasted time. In our case, many of those `adaptations' resulted only from the direct application of lessons learned during the project, that would be harder to recognise earlier in the project.
    \item Team members were involved in all aspects of the project, including development, verification and manufacturing. However, in the case of AI4Space it would be ideal to allocate at least manufacturing to another team or an external contractor, so as to not impede the momentum and progress of every iteration.
\end{itemize}

\begin{figure}[]
    \centering
    \includegraphics[width=\linewidth]{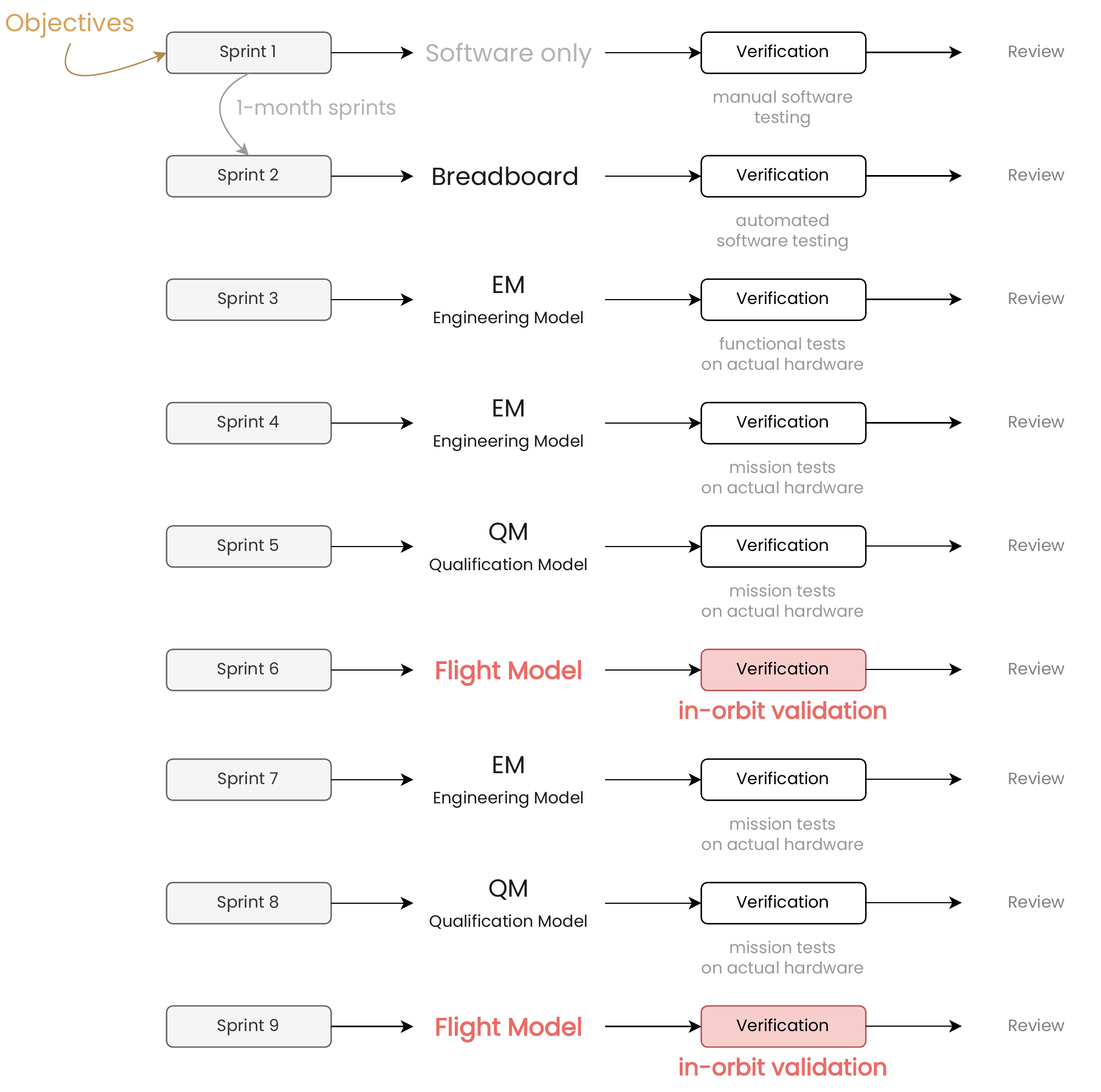}
    \caption{Theoretical example of AI4Space scrum schedule}
    \label{fig:theoretical_ai4space}
\end{figure}

\paragraph{}
The approach taken in the AI4Space fits the most basic approach of the triangle model, as described in \Cref{sec:the_triangle_model}. While the original goals of the mission were defined, after some iterations of design, proof of concepts and rudimentary testing, they were redefined so that the process could continue again; the team followed this cycle three times, back-tracking after the first breadboard proof-of-concept, and after the implementation of the AI algorithm.

As mentioned in \Cref{sec:caseintro}, we can theorise the effects of applying the ``Sat-Scrum'' method of \Cref{sec:sat_scrum} to AI4Space. This could lead to a more structured workflow, as shown in \Cref{fig:theoretical_ai4space}. In this case, two \emph{in-orbit validation} phases are implemented: The first is used as a technology demonstrator of the platform, only with some rudimentary functionality; the second would include a complete implementation of the original need. Some advantages of applying this method over a simple triangle would be:
\begin{itemize}[itemsep=0pt]
    \item Clearer project structure and task allocation, using well-defined procedures and tools, and streamlined interactions between developers
    \item ``Forced'' communication between members: encourages information sharing and reduces misunderstandings
    \item Verification and documentation would be required for each sprint. Documentation would only need to be updated every month, and not at very fast intervals that would render it obsolete immediately.
    \item The inclusion of more Engineering Models would prevent last-minute issues from blocking the production of Qualification Models, at the expense of some administrative overhead and cost.
\end{itemize}

However, we note that the structure of \Cref{fig:theoretical_ai4space} is not a `plan' in the conventional sense: Since Scrum is derived from empiricism and emergence, the actual decisions would be taken during the development cycle. Nonetheless, milestones such as launches cannot be changed, and have to be considered as invariants by the Scrum team.

\todo{Estimated manufacturing price?}
			
\subsection{ChipSat}
\begin{figure}[]
    \centering
    \includegraphics[width=\linewidth]{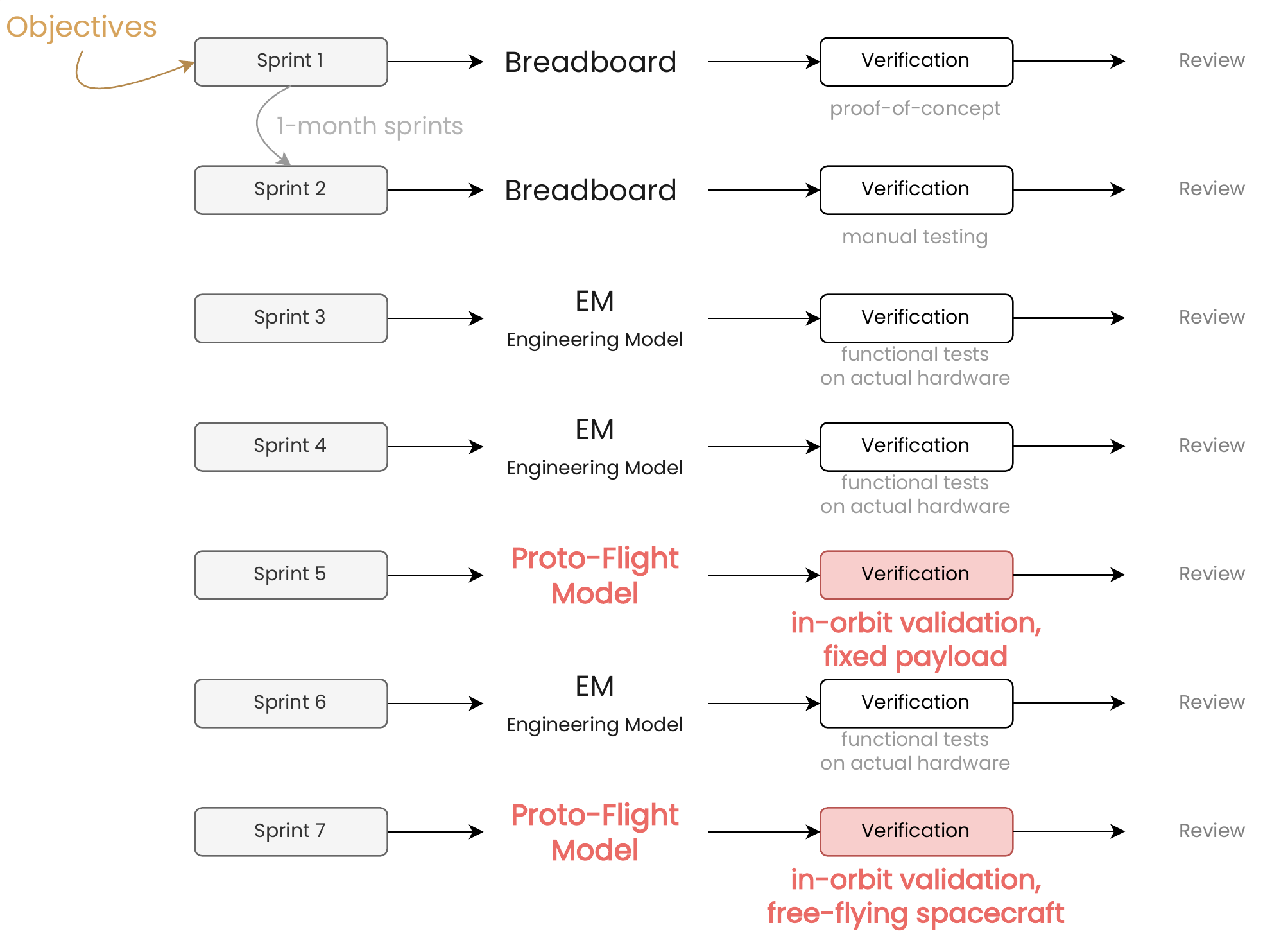}
    \caption{Theoretical example of the ChipSat scrum schedule}
    \label{fig:theoretical_chipsat}
\end{figure}

The ChipSat project also followed some agile principles, in combination with some traditional planning:
\begin{itemize}
    \item A \textbf{Gantt chart} was made using assumptions early in the project. However, it was seen that frequent updates to the Gantt chart were inevitable,  but it is regularly updated after any new development.
    \item Continuous changes in the requirements, capabilities and results meant that the project had to be \textbf{easily adaptable} to any changes. For example, the \acs{PCB} design had to be updated a few times in all stages of the project to accomodate new geometry restrictions.
    \item The team did not set in stone the number of iterations or designs that would be manufactured. Rather, since the procurement and manufacturing cost was low, a complete chipsat board was procured after every design iteration.
    \item Software and hardware development both started from day 0. This produced invaluable information regarding the feasibility of different design choices, and provided an approximate (yet adequate) reference model for internal use, but also for all external collaborators.
    \item \textbf{Project risk} was identified and attempted to be reduced. Especially regarding chip shortages, the procurement process for critical \acsp{IC} was initiated immediately after a relevant design decision is taken, to prevent chip shortages from delaying the project by many months.
    \item Due to uncertainties, different ``branches'' or options of the project had to be worked on at the same time until resolution (e.g.~version with/without batteries, with deployable/fixed antenna, etc.)
\end{itemize}

Modelling the design process with the triangle model of \Cref{sec:the_triangle_model} reveals that the definition of goals in this project (in terms of payload, features and performance) began after building the physical proof-of-concept --- i.e.~the model did not start from the top, but started from the implementation, and then back-tracked to the goals definition.

\Cref{fig:theoretical_chipsat} demonstrates a hypothetical Scrum approach of the ChipSat development. Here we note that in-orbit validation can first happen as an integrated payload (or even a high-altitude baloon flight), while the ChipSat can be released as an independent spacecraft on a second mission.

The advantages of this approach would be similar to the ones for \hyperref[sec:ai4spaceresults]{AI4Space}. However, the smaller size of the system, combined with a mass of \(<\)\SI{20}{\gram} for the freestanding board, would make it more convenient to produce multiple models and test in real-world conditions. These produced functional ChipSat models could even be used on the ground for outreach, or as wireless, self-sufficient Internet of Things (IoT) nodes \autocite{adams_theory_2020}.

\section{Conclusion}

In this paper, we explored how the unique characteristics of sub-CubeSat spacecraft can provide opportunities for tailored design methods, which take advantage of their low cost and potential for rapid manufacturing and testing. To develop a tailored model, we received inspiration from two case studies in the University of Luxembourg, and from the principles of Agile systems engineering. We then devised two lifecycle models for the development of sub-CubeSat spacecraft, and we verified their theoretical application to the two case studies.
We observed that by following one of the proposed models, we could keep development time for such missions down to 6--9 months, while iterating on physical builds of the system every \textasciitilde 1 month. This approach could reduce project risk by giving almost instant feedback to the designers about their design choices. By taking advantage of economy-of-scale effects, manufacturers can launch sub-CubeSat spacecraft as part of the development process, but keep manufacturing and testing costs in proportion to the spacecraft's mass.
For future work, we propose more research on how other agile frameworks tailored to systems engineering (such as Crystal, SAFe or RUP \autocite{boehm_balancing_2004,bott_analysis_2019}) and MBSE could be applied to gram-scale spacecraft. We also propose investigation or development of specific tools to aid teams in their day-to-day task management. Finally, we propose further validating proposed methods for these spacecraft, by comparing the application of different methods to actual case studies.







\AtNextBibliography{\footnotesize}
\printbibliography

@book{yin_case_2014,
	edition = {5},
	title = {Case Study Research: Design and Methods},
	isbn = {978-1-4522-4256-9},
	shorttitle = {Case Study Research},
	pagetotal = {313},
	publisher = {{SAGE} Publications},
	author = {Yin, Robert K.},
	date = {2014},
	langid = {english},
}

@report{beck_agile_2001,
	title = {The Agile Manifesto},
	url = {https://agilemanifesto.org/},
	author = {Beck, Kent and Beedle, Mike and van Bennekum, Arie and Cockburn, Alistair and Cunningham, Ward and Fowler, Martin and Martin, Robert C. and Mellor, Steve and Thomas, Dave and Grenning, James},
	urldate = {2022-08-30},
	date = {2001-02},
}

@inproceedings{barnhart_enabling_2007,
	location = {Logan, Utah},
	title = {Enabling Space Sensor Networks with {PCBSat}},
	volume = {{SSC}07-{IV}-4},
	url = {https://www.researchgate.net/publication/261297212_Enabling_Space_Sensor_Networks_with_PCBSat},
	eventtitle = {{AIAA}/{USU} Small Satellite Conference 2007},
	author = {Barnhart, David and Vladimirova, Tanya and Sweeting, Martin and Balthazor, Richard and Enloe, C. and Krause, L. and Lawrence, T. and Mcharg, M. and Lyke, J. and White, Jim and Baker, Adam},
	date = {2007-08-13},
}

@article{berthoud_university_2019,
	title = {University {CubeSat} Project Management for Success},
	url = {https://research-information.bris.ac.uk/en/publications/university-cubesat-project-management-for-success},
	journaltitle = {33rd Annual {AIAA}/{USUConference} on Small Satellites {SSC}19-{WKIII}-07},
	author = {Berthoud, Lucy and Swartwout, Michael and Cutler, James and Klumpar, David and Larsen, Jesper and Nielsen, Jens Dalsgaard},
	date = {2019-08-03},
	note = {Publisher: Utah State University},
}

@report{ECSS-E-ST-10C,
	title = {{ECSS}-E-{ST}-10C Rev.1 – System engineering general requirements},
	url = {https://ecss.nl/standard/ecss-e-st-10c-rev-1-system-engineering-general-requirements-15-february-2017/},
	institution = {European Space Agency},
	author = {{ECSS Secretariat}},
	urldate = {2022-07-07},
	date = {2017-02-15},
}

@report{PQ1,
	title = {The {PocketQube} Standard},
	url = {https://static1.squarespace.com/static/53d7dcdce4b07a1cdbbc08a4/t/5b34c395352f5303fcec6f45/1530184648111/PocketQube+Standard+issue+1+-+Published.pdf},
	institution = {Alba Orbital, {TU} Delft, G.A.U.S.S. Srl},
	author = {Radu, S and Uludag, M S and Speretta, S and Bouwmeester, J and Menicucci, A and Cervone, A and Dunn, A and Walkinshaw, T},
	urldate = {2022-08-30},
	date = {2018-06-07},
	langid = {english},
}

@article{peterson_when_2021,
	title = {When Worlds Collide --- A comparative analysis of issues impeding adoption of Agile for hardware},
	volume = {1},
	issn = {2732-527X},
	url = {http://www.cambridge.org/core/journals/proceedings-of-the-design-society/article/when-worlds-collide-a-comparative-analysis-of-issues-impeding-adoption-of-agile-for-hardware/6A4FF14A1D3313DE07D0330C9976CAE1},
	doi = {10.1017/pds.2021.606},
	pages = {3451--3460},
	journaltitle = {Proceedings of the Design Society},
	author = {Peterson, Matthew and Summers, Joshua},
	urldate = {2022-08-30},
	date = {2021-08},
	langid = {english},
	note = {Publisher: Cambridge University Press},
}

@online{north_whats_2007,
	title = {What's in a Story?},
	url = {https://dannorth.net/whats-in-a-story/},
	author = {North, Dan},
	urldate = {2022-09-14},
	date = {2007-02-11},
	langid = {english},
}

@inproceedings{boggero_mbse_2021,
	title = {An {MBSE} Architectural Framework for the Agile Definition of System Stakeholders, Needs and Requirements},
	pages = {3076},
	booktitle = {{AIAA} Aviation 2021 Forum},
	author = {Boggero, Luca and Ciampa, Pier Davide and Nagel, Björn},
	date = {2021},
}

@book{beck_test_2002,
	title = {Test Driven Development: By Example},
	isbn = {978-0-321-14653-3},
	url = {https://www.oreilly.com/library/view/test-driven-development/0321146530/},
	series = {Addison-Wesley Signature Series},
	shorttitle = {Test Driven Development},
	publisher = {Addison Wesley Professional, Pearson Education distributor},
	author = {Beck, Kent},
	urldate = {2022-08-31},
	date = {2002},
}

@article{dillon_faster-better-cheaper_2015,
	title = {Faster-Better-Cheaper Projects: Too Much Risk or Overreaction to Perceived Failure?},
	volume = {62},
	issn = {1558-0040},
	doi = {10.1109/TEM.2015.2404295},
	shorttitle = {Faster-Better-Cheaper Projects},
	pages = {141--149},
	number = {2},
	journaltitle = {{IEEE} Transactions on Engineering Management},
	author = {Dillon, Robin L. and Madsen, Peter M.},
	date = {2015-05},
}

@book{blessing_drm_2009,
	title = {{DRM}, a Design Research Methodology},
	isbn = {978-1-84882-587-1},
	url = {https://link.springer.com/book/10.1007/978-1-84882-587-1},
	pagetotal = {411},
	publisher = {Springer Science \& Business Media},
	author = {Blessing, Lucienne T. M. and Chakrabarti, Amaresh},
	date = {2009-06-13},
	langid = {english},
}

@report{nasa_cubesat_launch_initiative_cubesat_2017,
	title = {{CubeSat} 101: Basic Concepts and Processes for First-Time {CubeSat} Developers},
	url = {https://www.nasa.gov/sites/default/files/atoms/files/nasa_csli_cubesat_101_508.pdf},
	author = {{NASA CubeSat Launch Initiative}},
	date = {2017-10},
	langid = {english},
}

@thesis{sousa_cubesat_2021,
	title = {{CubeSat} Development Framework},
	url = {https://scholar.afit.edu/etd/4964},
	institution = {Air Force Institute of Technology},
	type = {phdthesis},
	author = {Sousa, William},
	date = {2021-03-01},
}

@article{bandecchi_concurrent_1999,
	title = {Concurrent engineering applied to space mission assessment and design},
	volume = {99},
	url = {https://www.esa.int/esapub/bulletin/bullet99/bande99.pdf},
	journaltitle = {{ESA} Bulletin},
	author = {Bandecchi, Massimo and Melton, B. and Ongaro, Franco},
	date = {1999-09},
}

@article{perez_survey_2016,
	title = {A Survey of Current Femtosatellite Designs, Technologies, and Mission Concepts},
	volume = {5},
	url = {https://ui.adsabs.harvard.edu/abs/2016JSSat...5..467P},
	pages = {467--482},
	journaltitle = {Journal of Small Satellites},
	author = {Perez, Tracie R. and Subbarao, Kamesh},
	urldate = {2022-09-09},
	date = {2016-10-01},
}

@software{robot_framework_foundation_robot_nodate,
	title = {Robot Framework},
	url = {https://robotframework.org/},
	author = {{Robot Framework Foundation}},
	urldate = {2022-09-14},
}

@software{bulski_construct_2020,
	title = {Construct},
	url = {https://construct.readthedocs.io/en/latest/},
	author = {Bulski, Arkadiusz and Filiba, Tomer and Simpson, Corbin},
	urldate = {2022-09-14},
	date = {2020-01},
}

@inproceedings{ciancarini_open_2020,
	location = {Cham},
	title = {An Open Source Environment for an Agile Development Model},
	isbn = {978-3-030-47240-5},
	doi = {10.1007/978-3-030-47240-5_15},
	series = {{IFIP} Advances in Information and Communication Technology},
	pages = {148--162},
	booktitle = {Open Source Systems},
	publisher = {Springer International Publishing},
	author = {Ciancarini, Paolo and Missiroli, Marcello and Poggi, Francesco and Russo, Daniel},
	editor = {Ivanov, Vladimir and Kruglov, Artem and Masyagin, Sergey and Sillitti, Alberto and Succi, Giancarlo},
	date = {2020},
	langid = {english},
}

@thesis{adams_theory_2020,
	title = {Theory and Applications of Gram-Scale Spacecraft},
	rights = {Attribution 4.0 International},
	url = {https://ecommons.cornell.edu/handle/1813/70469},
	institution = {Cornell University},
	type = {phdthesis},
	author = {Adams, Van Hunter},
	urldate = {2022-07-08},
	date = {2020-05},
	langid = {english},
	doi = {10.7298/b0bt-8v62},
}

@article{cappaert_building_2018,
	title = {Building, Deploying and Operating a Cubesat Constellation - Exploring the Less Obvious Reasons Space is Hard},
	url = {https://digitalcommons.usu.edu/smallsat/2018/all2018/274},
	journaltitle = {Small Satellite Conference},
	author = {Cappaert, Jeroen},
	date = {2018-08-07},
}

@collection{sebok,
	location = {Hoboken, {NJ}},
	edition = {v. 2.6},
	title = {The Guide to the Systems Engineering Body of Knowledge ({SEBoK})},
	url = {https://www.sebokwiki.org/wiki/System_Lifecycle_Process_Models:_Incremental},
	editor = {{SEBoK Editorial Board}},
	urldate = {2022-09-09},
	date = {2022-05-20},
}

@inproceedings{darrin_agile_2017,
	title = {The Agile Manifesto, design thinking and systems engineering},
	doi = {10.1109/SYSCON.2017.7934765},
	eventtitle = {2017 Annual {IEEE} International Systems Conference ({SysCon})},
	pages = {1--5},
	author = {Darrin, M. Ann Garrison and Devereux, William S.},
	date = {2017-04},
	note = {{ISSN}: 2472-9647},
}

@book{boehm_balancing_2004,
	location = {Boston},
	title = {Balancing agility and discipline: a guide for the perplexed},
	isbn = {978-0-321-62388-1},
	url = {https://dl.acm.org/doi/10.5555/861419},
	shorttitle = {Balancing agility and discipline},
	publisher = {Addison-Wesley},
	author = {Boehm, Barry W.},
	editora = {Turner, Richard},
	editoratype = {collaborator},
	urldate = {2022-08-29},
	date = {2004},
}

@article{carson_421_2013,
	title = {4.2.1 Can Systems Engineering be Agile? Development Lifecycles for Systems, Hardware, and Software},
	volume = {23},
	issn = {2334-5837},
	url = {http://onlinelibrary.wiley.com/doi/abs/10.1002/j.2334-5837.2013.tb03001.x},
	doi = {10.1002/j.2334-5837.2013.tb03001.x},
	shorttitle = {4.2.1 Can Systems Engineering be Agile?},
	pages = {16--28},
	number = {1},
	journaltitle = {{INCOSE} International Symposium},
	author = {Carson, Ronald S.},
	urldate = {2022-08-30},
	date = {2013},
	langid = {english},
}

@article{yassine_information_2003,
	title = {Information hiding in product development: the design churn effect},
	volume = {14},
	issn = {1435-6066},
	url = {https://doi.org/10.1007/s00163-003-0036-2},
	doi = {10.1007/s00163-003-0036-2},
	shorttitle = {Information hiding in product development},
	pages = {145--161},
	number = {3},
	journaltitle = {Research in Engineering Design},
	shortjournal = {Res Eng Design},
	author = {Yassine, Ali and Joglekar, Nitin and Braha, Dan and Eppinger, Steven and Whitney, Daniel},
	urldate = {2022-09-12},
	date = {2003-11-01},
	langid = {english},
}

@report{bundesrepublik_deutschland_v-modell_2006,
	title = {V-Modell {XT}},
	url = {http://ftp.uni-kl.de/pub/v-modell-xt/Release-1.1-eng/Dokumentation/pdf/V-Modell-XT-eng-Teil1.pdf},
	author = {{Bundesrepublik Deutschland}},
	urldate = {2022-09-07},
	date = {2006-01-31},
}

@video{a_s_u_news_suncube_2016,
	title = {{SunCube} miniature satellites},
	url = {https://vimeo.com/161656838},
	author = {{A. S. U. News}},
	urldate = {2022-09-09},
	date = {2016-04-05},
}

@inproceedings{lubian-arenillas_nanosatellite_2019,
	location = {Madrid},
	title = {Nanosatellite development methodology and preliminary design guides for the {NANOSTAR} Project},
	rights = {https://creativecommons.org/licenses/by-nc-nd/3.0/es/},
	url = {https://www.eucass.eu/doi/EUCASS2019-0562.pdf},
	eventtitle = {European Conference for Aeronautics and Space Sciences ({EUCASS} 2019)},
	pages = {1--11},
	publisher = {E.T.S. de Ingeniería Aeronáutica y del Espacio ({UPM})},
	author = {Lubián-Arenillas, Daniel and Alvarez Romero, Jose Miguel and Bermejo Ballesteros, Juan and García, Sergio and Cubas Cano, Javier and Roibás-Millán, Elena},
	urldate = {2022-09-08},
	date = {2019-07},
}

@article{wolf_multiprocessor_2008,
	title = {Multiprocessor System-on-Chip ({MPSoC}) Technology},
	volume = {27},
	issn = {1937-4151},
	doi = {10.1109/TCAD.2008.923415},
	pages = {1701--1713},
	number = {10},
	journaltitle = {{IEEE} Transactions on Computer-Aided Design of Integrated Circuits and Systems},
	author = {Wolf, Wayne and Jerraya, Ahmed Amine and Martin, Grant},
	date = {2008-10},
}

@article{honore-livermore_agile_2021,
	title = {An Agile Systems Engineering Analysis of a University {CubeSat} Project Organization},
	volume = {31},
	issn = {2334-5837},
	url = {http://onlinelibrary.wiley.com/doi/abs/10.1002/j.2334-5837.2021.00904.x},
	doi = {10.1002/j.2334-5837.2021.00904.x},
	pages = {1334--1348},
	number = {1},
	journaltitle = {{INCOSE} International Symposium},
	author = {Honoré-Livermore, Evelyn and Lyells, Ron and Garrett, Joseph L. and Angier, Rock and Epps, Bob},
	urldate = {2022-09-08},
	date = {2021},
	langid = {english},
}

@inproceedings{lill_agile_2018,
	title = {Agile Mission Operations in the {CubeSat} Project {MOVE}-{II}},
	url = {https://arc.aiaa.org/doi/abs/10.2514/6.2018-2635},
	doi = {10.2514/6.2018-2635},
	booktitle = {2018 {SpaceOps} conference},
	author = {Lill, Alexander and Zwickl, Thomas and Costescu, Constantin and Patzwahl, Lucie and Soare, Cristian and Langer, Martin},
	date = {2018-05-25},
}

@article{labarge_cubesat_2014,
	title = {{CubeSat} – An Agile System Architecture?},
	volume = {17},
	issn = {2156-4868},
	url = {http://onlinelibrary.wiley.com/doi/abs/10.1002/inst.201417227},
	doi = {10.1002/inst.201417227},
	pages = {27--30},
	number = {2},
	journaltitle = {{INSIGHT}},
	author = {{LaBarge}, Ralph},
	urldate = {2022-09-09},
	date = {2014},
	langid = {english},
}

@inproceedings{coyle_eecsat_2020,
	location = {West Point, New York, {USA}},
	title = {{EECSat}: {CubeSat} Development},
	url = {http://www.ieworldconference.org/content/WP2020/Papers/GDRKMCC_20_32.pdf},
	shorttitle = {{EECSat}},
	eventtitle = {2020 Annual General Donald R. Keith Memorial Capstone Conference},
	pages = {6},
	author = {Coyle, Sean and Burk, Roger and Kedrowitsch, Alexander},
	date = {2020-04-30},
	langid = {english},
}

@article{kiesbye_hardware---loop_2019,
	title = {Hardware-In-The-Loop and Software-In-The-Loop Testing of the {MOVE}-{II} {CubeSat}},
	volume = {6},
	rights = {http://creativecommons.org/licenses/by/3.0/},
	issn = {2226-4310},
	url = {https://www.mdpi.com/2226-4310/6/12/130},
	doi = {10.3390/aerospace6120130},
	pages = {130},
	number = {12},
	journaltitle = {Aerospace},
	author = {Kiesbye, Jonis and Messmann, David and Preisinger, Maximilian and Reina, Gonzalo and Nagy, Daniel and Schummer, Florian and Mostad, Martin and Kale, Tejas and Langer, Martin},
	urldate = {2022-09-09},
	date = {2019-12},
	langid = {english},
	note = {Number: 12
Publisher: Multidisciplinary Digital Publishing Institute},
}

@online{tyvak_trestles_2021,
	title = {Trestles 6U/12U Platform Specification Booklet},
	url = {https://rsdo.gsfc.nasa.gov/images/catalog-rapidIV/Tyvak_Trestles_6U-12U_Catalog_Brochure.pdf},
	author = {{Tyvak}},
	urldate = {2022-09-08},
	date = {2021-01-19},
}

@inproceedings{royce_managing_1970,
	location = {Los Angeles, {USA}},
	title = {Managing the Development of Large Software Systems},
	url = {https://dl.acm.org/doi/10.5555/41765.41801},
	booktitle = {Technical Papers of Western Electronic Show and Convention},
	author = {Royce, Winston W.},
	date = {1970-08},
}

@article{dingsoyr_decade_2012,
	title = {A decade of agile methodologies: Towards explaining agile software development},
	volume = {85},
	issn = {0164-1212},
	url = {https://www.sciencedirect.com/science/article/pii/S0164121212000532},
	doi = {10.1016/j.jss.2012.02.033},
	series = {Special Issue: Agile Development},
	shorttitle = {A decade of agile methodologies},
	pages = {1213--1221},
	number = {6},
	journaltitle = {Journal of Systems and Software},
	shortjournal = {Journal of Systems and Software},
	author = {Dingsøyr, Torgeir and Nerur, Sridhar and Balijepally, {VenuGopal} and Moe, Nils Brede},
	urldate = {2022-09-09},
	date = {2012-06-01},
	langid = {english},
}

@article{gong_design_2022,
	title = {Design of foldable {PCBSat} enabling three-axis attitude control},
	volume = {192},
	issn = {0094-5765},
	url = {https://www.sciencedirect.com/science/article/pii/S0094576521006391},
	doi = {10.1016/j.actaastro.2021.12.004},
	pages = {291--300},
	journaltitle = {Acta Astronautica},
	shortjournal = {Acta Astronautica},
	author = {Gong, Haoran and Gong, Shengping},
	urldate = {2022-09-09},
	date = {2022-03-01},
	langid = {english},
}

@online{noauthor_fossa_2021,
	title = {{FOSSA} Systems - Our dedicated picosatellite platforms for {IoT}},
	url = {https://fossa.systems/satellites/},
	urldate = {2022-09-09},
	date = {2021-05-27},
	langid = {american},
}

@online{noauthor_unicorn_nodate,
	title = {Unicorn 2 Platform},
	url = {http://www.albaorbital.com/unicorn-2},
	titleaddon = {Alba Orbital},
	urldate = {2022-09-09},
	langid = {british},
}

@inproceedings{gangestad_flight_2015,
	title = {Flight Results from {AeroCube}-6: A Radiation Dosimeter Mission in the 0.5 U Form Factor},
	shorttitle = {Flight Results from {AeroCube}-6},
	booktitle = {12th Annual {CubeSat} Developers Workshop, San Luis Obispo, {CA}},
	author = {Gangestad, Joseph and Rowen, Darren and Hardy, Brian and Coffman, Christopher and O'Brien, Paul},
	date = {2015},
}

@inproceedings{speretta_cubesats_2016,
	title = {{CubeSats} to {PocketQubes}: Opportunities and Challenges},
	url = {https://www.researchgate.net/publication/308777632_CubeSats_to_PocketQubes_Opportunities_and_Challenges},
	shorttitle = {{CubeSats} to {PocketQubes}},
	eventtitle = {67th International Astronautical Congress},
	author = {Speretta, Stefano and Soriano, Tatiana and Bouwmeester, Jasper and Carvajal-Godínez, Johan and Menicucci, Alessandra and Watts, Trevor and Sundaramoorthy, Prem and Guo, Jian and Gill, Eberhard},
	date = {2016-09-26},
}

@article{saeed_cubesat_2020,
	title = {{CubeSat} Communications: Recent Advances and Future Challenges},
	volume = {22},
	issn = {1553-877X},
	doi = {10.1109/COMST.2020.2990499},
	shorttitle = {{CubeSat} Communications},
	pages = {1839--1862},
	number = {3},
	journaltitle = {{IEEE} Communications Surveys \& Tutorials},
	author = {Saeed, Nasir and Elzanaty, Ahmed and Almorad, Heba and Dahrouj, Hayssam and Al-Naffouri, Tareq Y. and Alouini, Mohamed-Slim},
	date = {2020},
	note = {Conference Name: {IEEE} Communications Surveys \& Tutorials},
}

@inproceedings{angeli_paving_2014,
	location = {Montréal, Quebec, Canada},
	title = {The Paving Stones: initial feed-back on an attempt to apply the {AGILE} principles for the development of a {CubeSat} space mission to Mars},
	url = {http://proceedings.spiedigitallibrary.org/proceeding.aspx?doi=10.1117/12.2056377},
	doi = {10.1117/12.2056377},
	shorttitle = {The Paving Stones},
	eventtitle = {{SPIE} Astronomical Telescopes + Instrumentation},
	pages = {91500W},
	author = {Segret, Boris and Semery, Alain and Vannitsen, Jordan and Mosser, Benoît and Miau, Jiun-Jih and Juang, Jyh-Ching and Deleflie, Florent},
	editor = {Angeli, George Z. and Dierickx, Philippe},
	urldate = {2022-09-08},
	date = {2014-08-04},
}

@thesis{decker_systems-engineering_2016,
	title = {A systems-engineering assessment of multiple {CubeSat} build approaches},
	rights = {M.I.T. theses are protected by copyright. They may be viewed from this source for any purpose, but reproduction or distribution in any format is prohibited without written permission. See provided {URL} for inquiries about permission.},
	url = {https://dspace.mit.edu/handle/1721.1/105560},
	institution = {Massachusetts Institute of Technology},
	type = {Thesis},
	author = {Decker, Zachary Scott},
	urldate = {2022-09-08},
	date = {2016},
	note = {Accepted: 2016-12-05T19:10:36Z},
}

@article{kohlbacher_agile_2011,
	title = {Do agile software development practices increase customer satisfaction in Systems Engineering projects?},
	doi = {10.1109/SYSCON.2011.5929091},
	author = {Kohlbacher, Markus and Stelzmann, Ernst and Maierhofer, Sabine},
	date = {2011-04-01},
}

@article{bott_analysis_2019,
	title = {An Analysis of Theories Supporting Agile Scrum and the Use of Scrum in Systems Engineering},
	volume = {32},
	doi = {10.1080/10429247.2019.1659701},
	pages = {1--10},
	journaltitle = {Engineering Management Journal},
	shortjournal = {Engineering Management Journal},
	author = {Bott, Mitch and Mesmer, Bryan},
	date = {2019-09-20},
}

@article{hein_evaluating_2020,
	title = {Evaluating engineering design methods: taking inspiration from software engineering and the health sciences},
	volume = {1},
	issn = {2633-7762},
	url = {http://www.cambridge.org/core/journals/proceedings-of-the-design-society-design-conference/article/pevaluating-engineering-design-methods-taking-inspiration-from-software-engineering-and-the-health-sciences/48034F035D9E112F46FC5D38C8A0FADA},
	doi = {10.1017/dsd.2020.317},
	shorttitle = {Evaluating engineering design methods},
	pages = {1901--1910},
	journaltitle = {Proceedings of the Design Society: {DESIGN} Conference},
	author = {Hein, A. M. and Lamé, G.},
	urldate = {2022-09-07},
	date = {2020-05},
	langid = {english},
	note = {Publisher: Cambridge University Press},
}

@thesis{manchester_centimeter-scale_2015,
	title = {Centimeter-Scale Spacecraft: Design, Fabrication, And Deployment},
	url = {https://ecommons.cornell.edu/handle/1813/41055},
	shorttitle = {Centimeter-Scale Spacecraft},
	type = {phdthesis},
	author = {Manchester, Zachary},
	urldate = {2022-02-17},
	date = {2015-08-17},
	langid = {american},
}

@inproceedings{garzaniti_effectiveness_2019,
	title = {Effectiveness of the Scrum Methodology for Agile Development of Space Hardware},
	doi = {10.1109/AERO.2019.8741892},
	eventtitle = {2019 {IEEE} Aerospace Conference},
	pages = {1--8},
	author = {Garzaniti, Nicola and Briatore, Simone and Fortin, Clément and Golkar, Alessandro},
	date = {2019-03},
	note = {{ISSN}: 1095-323X},
}

@inproceedings{tang_mbse_2018,
	title = {An {MBSE} framework to support agile functional definition of an avionics system},
	url = {https://link.springer.com/chapter/10.1007/978-3-030-04209-7_14},
	doi = {10.1007/978-3-030-04209-7_14},
	pages = {168--178},
	booktitle = {International Conference on Complex Systems Design \& Management},
	publisher = {Springer},
	author = {Tang, Jian and Zhu, Shaofan and Faudou, Raphaël and Gauthier, Jean-Marie},
	date = {2018},
}

@article{alanazi_engineering_2019,
	title = {Engineering Methodology for Student-Driven {CubeSats}},
	volume = {6},
	rights = {http://creativecommons.org/licenses/by/3.0/},
	issn = {2226-4310},
	url = {https://www.mdpi.com/2226-4310/6/5/54},
	doi = {10.3390/aerospace6050054},
	pages = {54},
	number = {5},
	journaltitle = {Aerospace},
	author = {Alanazi, Abdulaziz and Straub, Jeremy},
	urldate = {2022-09-06},
	date = {2019-05},
	langid = {english},
	note = {Number: 5
Publisher: Multidisciplinary Digital Publishing Institute},
}

@article{nieto-peroy_cubesat_2019,
	title = {{CubeSat} Mission: From Design to Operation},
	volume = {9},
	rights = {http://creativecommons.org/licenses/by/3.0/},
	issn = {2076-3417},
	url = {https://www.mdpi.com/2076-3417/9/15/3110},
	doi = {10.3390/app9153110},
	shorttitle = {{CubeSat} Mission},
	pages = {3110},
	number = {15},
	journaltitle = {Applied Sciences},
	author = {Nieto-Peroy, Cristóbal and Emami, M. Reza},
	urldate = {2022-09-06},
	date = {2019-01},
	langid = {english},
	note = {Number: 15
Publisher: Multidisciplinary Digital Publishing Institute},
}

@inproceedings{carpenter_is_2014,
	title = {Is Agile Too Fragile for Space-Based Systems Engineering?},
	doi = {10.1109/SMC-IT.2014.13},
	eventtitle = {2014 {IEEE} International Conference on Space Mission Challenges for Information Technology},
	pages = {38--45},
	author = {Carpenter, Scott E. and Dagnino, Aldo},
	date = {2014-09},
}

@book{douglass_agile_2015,
	edition = {1st edition.},
	title = {Agile Systems Engineering},
	isbn = {978-0-12-802349-5},
	url = {https://www.oreilly.com/library/view/agile-systems-engineering/9780128023495/},
	publisher = {Morgan Kaufmann},
	author = {Douglass, Bruce},
	urldate = {2022-02-17},
	date = {2015},
}

@book{beck_extreme_2008,
	location = {Boston},
	edition = {2nd ed.},
	title = {Extreme programming explained: embrace change},
	isbn = {978-0-321-27865-4},
	url = {https://www.oreilly.com/library/view/extreme-programming-explained/0321278658/},
	series = {The {XP} series},
	shorttitle = {Extreme programming explained},
	publisher = {Addison-Wesley Professional},
	author = {Beck, Kent},
	editora = {Andres, Cynthia},
	editoratype = {collaborator},
	urldate = {2022-08-31},
	date = {2008},
}

@inproceedings{kasauli_safety-critical_2018,
	title = {Safety-Critical Systems and Agile Development: A Mapping Study},
	doi = {10.1109/SEAA.2018.00082},
	shorttitle = {Safety-Critical Systems and Agile Development},
	eventtitle = {2018 44th Euromicro Conference on Software Engineering and Advanced Applications ({SEAA})},
	pages = {470--477},
	author = {Kasauli, Rashidah and Knauss, Eric and Kanagwa, Benjamin and Nilsson, Agneta and Calikli, Gul},
	date = {2018-08},
}

@article{sweeting_modern_2018,
	title = {Modern Small Satellites-Changing the Economics of Space},
	volume = {106},
	issn = {1558-2256},
	doi = {10.1109/JPROC.2018.2806218},
	pages = {343--361},
	number = {3},
	journaltitle = {Proceedings of the {IEEE}},
	author = {Sweeting, Martin N.},
	date = {2018-03},
}

@inproceedings{clark_system_2009,
	title = {System of Systems Engineering and Family of Systems Engineering from a standards, V-Model, and Dual-V Model perspective},
	doi = {10.1109/SYSTEMS.2009.4815831},
	eventtitle = {2009 3rd Annual {IEEE} Systems Conference},
	pages = {381--387},
	booktitle = {2009 3rd Annual {IEEE} Systems Conference},
	author = {Clark, John O.},
	date = {2009-03},
}

@misc{hein_attosats_2019,
	title = {{AttoSats}: {ChipSats}, other Gram-Scale Spacecraft, and Beyond},
	url = {http://arxiv.org/abs/1910.12559},
	doi = {10.48550/arXiv.1910.12559},
	shorttitle = {{AttoSats}},
	publisher = {{arXiv}},
	author = {Hein, Andreas M. and Burkhardt, Zachary and Eubanks, T. Marshall},
	urldate = {2022-08-31},
	date = {2019-12-31},
	eprinttype = {arxiv},
	eprint = {1910.12559 [astro-ph, physics:physics]},
}

@report{schwaver_definitive_2020,
	title = {The Definitive Guide to Scrum: The Rules of the Game},
	url = {https://scrumguides.org/docs/scrumguide/v2020/2020-Scrum-Guide-US.pdf},
	shorttitle = {The Scrum Guide},
	author = {Schwaver, Ken and Sutherland, Jeff},
	date = {2020-11-09},
}

@online{de_vos_documentation_2022,
	title = {Documentation of Open Hardware},
	url = {https://hackmd.io/@Oggo2XIlRZ6wwlsXi_vc8Q/By3DNodtq},
	titleaddon = {Delft Open Hardware Academy},
	author = {de Vos, Jerry and Urra, Jose},
	urldate = {2022-08-30},
	date = {2022-08-15},
	langid = {american},
}

@article{shahin_continuous_2017,
	title = {Continuous Integration, Delivery and Deployment: A Systematic Review on Approaches, Tools, Challenges and Practices},
	volume = {5},
	issn = {2169-3536},
	doi = {10.1109/ACCESS.2017.2685629},
	shorttitle = {Continuous Integration, Delivery and Deployment},
	pages = {3909--3943},
	journaltitle = {{IEEE} Access},
	author = {Shahin, Mojtaba and Ali Babar, Muhammad and Zhu, Liming},
	date = {2017},
	note = {Conference Name: {IEEE} Access},
}

@report{ECSS-E-HB-10-02A,
	title = {{ECSS}-E-{HB}-10-02A – Verification guidelines},
	url = {https://ecss.nl/hbstms/ecss-e-10-02a-verification-guidelines/},
	institution = {European Space Agency},
	author = {{ECSS Secretariat}},
	urldate = {2022-08-31},
	date = {2020-12-17},
}

@article{fritz_hardware---loop_2015,
	title = {Hardware-in-the-loop environment for verification of a small satellite's on-board software},
	volume = {47},
	issn = {1270-9638},
	url = {https://www.sciencedirect.com/science/article/pii/S1270963815002783},
	doi = {10.1016/j.ast.2015.09.020},
	pages = {388--395},
	journaltitle = {Aerospace Science and Technology},
	shortjournal = {Aerospace Science and Technology},
	author = {Fritz, Michael and Winter, Sebastian and Freund, Juergen and Pflueger, Stefan and Zeile, Oliver and Eickhoff, Jens and Roeser, Hans-Peter},
	urldate = {2022-08-31},
	date = {2015-12-01},
	langid = {english},
}

@article{barnhart_low-cost_2009,
	title = {A low-cost femtosatellite to enable distributed space missions},
	volume = {64},
	issn = {0094-5765},
	url = {https://www.sciencedirect.com/science/article/pii/S0094576509000198},
	doi = {10.1016/j.actaastro.2009.01.025},
	pages = {1123--1143},
	number = {11},
	journaltitle = {Acta Astronautica},
	shortjournal = {Acta Astronautica},
	author = {Barnhart, David J. and Vladimirova, Tanya and Baker, Adam M. and Sweeting, Martin N.},
	urldate = {2022-08-30},
	date = {2009-06-01},
	langid = {english},
}

@online{triantafyllopoulou_qubik_2020,
	title = {The {QUBIK} Project: Ready for orbit},
	url = {https://libre.space/2020/11/19/the-qubik-project-ready-for-orbit/},
	shorttitle = {The {QUBIK} Project},
	titleaddon = {Libre Space Foundation},
	author = {Triantafyllopoulou, Nikoletta},
	urldate = {2022-08-30},
	date = {2020-11-19},
	langid = {american},
}

@report{PQ91,
	title = {{PQ}9 and {CS}14 Electrical and Mechanical Subsystem Interface Standard for {PocketQubes} and {CubeSats}},
	url = {https://dataverse.nl/dataset.xhtml?persistentId=doi:10.34894/6MVBCZ},
	institution = {{DataverseNL}},
	author = {Bouwmeester, Jasper and Chia, J and Boerci, M and Radu, S},
	urldate = {2022-08-30},
	date = {2018-07-04},
	langid = {english},
	doi = {10.34894/6MVBCZ},
}

@incollection{garzaniti_toward_2020,
	title = {Toward a Hybrid Agile Product Development Process},
	isbn = {978-3-030-42249-3},
	pages = {191--200},
	author = {Garzaniti, Nicola and Fortin, Clément and Golkar, Alessandro},
	date = {2020-02-01},
	doi = {10.1007/978-3-030-42250-9_18},
}

@article{dallmann_agile_2015,
	title = {An Agile Space Paradigm and the Prometheus {CubeSat} System},
	url = {https://digitalcommons.usu.edu/smallsat/2015/all2015/33},
	journaltitle = {Small Satellite Conference},
	author = {Dallmann, Nicholas and Delapp, Jerry and Enemark, Donald and Fairbanks, Thomas and Fortgang, Clifford and Guenther, David and Judd, Stephen and Kestell, Gayle and Lake, James and Martinez, John and {McCabe}, Kevin and Michel, John and Palmer, Joseph and Powell, Mitchell and Prichard, Dean and Proicou, Michael and Quinn, Heather and Reid, Robert and Schaller, Edward and Seitz, Daniel and Stein, Paul and Storms, Steven and Sullivan, Erica and Tripp, Justin and Warniment, Adam and Wheat, Robert},
	date = {2015-08-11},
}

@article{fischer_implementing_2017,
	title = {Implementing model-based system engineering for the whole lifecycle of a spacecraft},
	volume = {9},
	issn = {1868-2510},
	url = {https://doi.org/10.1007/s12567-017-0166-4},
	doi = {10.1007/s12567-017-0166-4},
	pages = {351--365},
	number = {3},
	journaltitle = {{CEAS} Space Journal},
	shortjournal = {{CEAS} Space J},
	author = {Fischer, P. M. and Lüdtke, D. and Lange, C. and Roshani, F.-C. and Dannemann, F. and Gerndt, A.},
	urldate = {2022-08-30},
	date = {2017-09-01},
	langid = {english},
}

@online{nasa_engineering__safety_center_aligning_2018,
	title = {Aligning System Development Models with Insight Approaches},
	url = {https://llis.nasa.gov/lesson/24502},
	titleaddon = {{NASA} Lessons Learned},
	author = {{NASA Engineering \& Safety Center}},
	urldate = {2022-08-30},
	date = {2018-08-23},
}

@article{HEEAGER201822,
	title = {A conceptual model of agile software development in a safety-critical context: A systematic literature review},
	volume = {103},
	issn = {0950-5849},
	url = {https://www.sciencedirect.com/science/article/pii/S0950584918301125},
	doi = {https://doi.org/10.1016/j.infsof.2018.06.004},
	pages = {22--39},
	journaltitle = {Information and Software Technology},
	author = {Heeager, Lise Tordrup and Nielsen, Peter Axel},
	date = {2018},
}

@online{abate_inexpensive_2019,
	title = {Inexpensive chip-size satellites orbit Earth},
	url = {https://news.stanford.edu/2019/06/03/chip-size-satellites-orbit-earth/},
	titleaddon = {Stanford News},
	author = {Abate, Tom},
	editora = {Manchester, Zachary},
	editoratype = {collaborator},
	urldate = {2022-08-29},
	date = {2019-06-03},
	langid = {english},
	note = {Section: Science \& Technology},
}

@article{manchester_kicksat_2013,
	title = {{KickSat}: A Crowd-Funded Mission to Demonstrate the World’s Smallest Spacecraft},
	url = {https://digitalcommons.usu.edu/smallsat/2013/all2013/111},
	shorttitle = {{KickSat}},
	journaltitle = {Small Satellite Conference},
	author = {Manchester, Zachary and Peck, Mason and Filo, Andrew},
	date = {2013-08-14},
}

@article{twiggs_thinsat_2018,
	title = {The {ThinSat} Program: Flight Opportunities for Education, Research and Industry},
	url = {https://digitalcommons.usu.edu/smallsat/2018/all2018/475},
	shorttitle = {The {ThinSat} Program},
	journaltitle = {Small Satellite Conference},
	author = {Twiggs, Robert and Zucherman, Aaron and Bujold, Emily and Counts, Nick and Colman, Christopher and Garcia, Jose and Diddle, Andrew and Zhirkina, Polina},
	date = {2018-08-06},
}

@report{CDS14,
	title = {{CubeSat} Design Specification Rev. 14.1},
	url = {https://www.cubesat.org/s/CDS-REV14_1-2022-02-09.pdf},
	number = {{CP}-{CDS}-R14.1},
	author = {{California Polytechnic State University}},
	urldate = {2022-08-26},
	date = {2022-02},
}

@online{noauthor_status_nodate,
	title = {Status Quo (Scaled) Agile 2020},
	url = {http://www.process-and-project.net/studien/studienunterseiten/status-quo-scaled-agile-2020-en/},
	titleaddon = {Process-and-Project.net},
	urldate = {2022-07-25},
	langid = {german},
}

@online{noauthor_15th_nodate,
	title = {15th Annual State Of Agile Report {\textbar} Digital.ai},
	url = {https://digital.ai/resource-center/analyst-reports/state-of-agile-report},
	urldate = {2022-07-25},
	langid = {english},
}

@misc{bychkov_using_2018,
	title = {Using Binary File Format Description Languages for Documenting, Parsing, and Verifying Raw Data in {TAIGA} Experiment},
	url = {http://arxiv.org/abs/1812.01324},
	doi = {10.48550/arXiv.1812.01324},
	publisher = {{arXiv}},
	author = {Bychkov, I. and Demichev, A. and Dubenskaya, J. and Fedorov, O. and Hmelnov, A. and Kazarina, Y. and Korosteleva, E. and Kostunin, D. and Kryukov, A. and Mikhailov, A. and Nguyen, M. D. and Polyakov, S. and Postnikov, E. and Shigarov, A. and Shipilov, D. and Zhurov, D.},
	urldate = {2022-07-11},
	date = {2018-12-04},
	eprinttype = {arxiv},
	eprint = {1812.01324 [astro-ph]},
}

@inproceedings{perrotin_taste_2012,
	location = {Berlin, Heidelberg},
	title = {{TASTE}: A Real-Time Software Engineering Tool-Chain Overview, Status, and Future},
	isbn = {978-3-642-25264-8},
	doi = {10.1007/978-3-642-25264-8_4},
	series = {Lecture Notes in Computer Science},
	shorttitle = {{TASTE}},
	pages = {26--37},
	booktitle = {{SDL} 2011: Integrating System and Software Modeling},
	publisher = {Springer},
	author = {Perrotin, Maxime and Conquet, Eric and Delange, Julien and Schiele, André and Tsiodras, Thanassis},
	editor = {Ober, Iulian and Ober, Ileana},
	date = {2012},
	langid = {english},
}

@article{haberfellner_agile_2005,
	title = {Agile Systems-Engineering versus Agile-Systems Engineering},
	volume = {15},
	doi = {10.1002/j.2334-5837.2005.tb00762.x},
	journaltitle = {{INCOSE} International Symposium},
	shortjournal = {{INCOSE} International Symposium},
	author = {Haberfellner, Reinhard and de Weck, Olivier},
	date = {2005-07-01},
}

@online{shea_nasa_2017,
	title = {{NASA} Systems Engineering Handbook Revision 2},
	url = {http://www.nasa.gov/connect/ebooks/nasa-systems-engineering-handbook},
	titleaddon = {{NASA}},
	type = {Text},
	author = {Shea, Garrett},
	urldate = {2022-07-05},
	date = {2017-06-20},
}

@inproceedings{gerrike_what_2017,
	location = {Vancouver, Canada},
	title = {What do we need to say about a design method?},
	url = {http://oro.open.ac.uk/50445/},
	eventtitle = {21th International Conference on Engineering Design ({ICED} 2015)},
	author = {Gerrike, Kilian and Eckert, Claudia and Stacey, Martin},
	urldate = {2022-02-22},
	date = {2017},
	langid = {english},
}

@misc{MAIVP,
	title = {{AcubeSAT} Manufacturing, Assembly, Integration and Verification File},
	url = {https://gitlab.com/acubesat/documentation/cdr-public/-/blob/master/MAIVP%20file/MAIVP.pdf},
	author = {{AcubeSAT Team}},
	date = {2021-05-17},
}

@online{kulu_nanosats_2021,
	title = {Nanosats Database},
	url = {https://www.nanosats.eu/index.html},
	titleaddon = {Nanosats Database},
	author = {Kulu, Erik},
	urldate = {2021-06-29},
	date = {2021-04-04},
	langid = {english},
}

@article{poghosyan_cubesat_2017,
	title = {{CubeSat} evolution: Analyzing {CubeSat} capabilities for conducting science missions},
	volume = {88},
	issn = {0376-0421},
	url = {https://www.sciencedirect.com/science/article/pii/S0376042116300951},
	doi = {10.1016/j.paerosci.2016.11.002},
	shorttitle = {{CubeSat} evolution},
	pages = {59--83},
	journaltitle = {Progress in Aerospace Sciences},
	shortjournal = {Progress in Aerospace Sciences},
	author = {Poghosyan, Armen and Golkar, Alessandro},
	urldate = {2021-06-24},
	date = {2017-01-01},
	langid = {english},
}

@book{aguirre_introduction_2013,
	location = {New York},
	title = {Introduction to Space Systems: Design and Synthesis},
	isbn = {978-1-4614-3757-4},
	url = {https://www.springer.com/gp/book/9781461437574},
	series = {Space Technology Library},
	shorttitle = {Introduction to Space Systems},
	publisher = {Springer-Verlag},
	author = {Aguirre, Miguel A.},
	urldate = {2021-05-23},
	date = {2013},
	langid = {english},
	doi = {10.1007/978-1-4614-3758-1},
}

@article{faure_toward_2017,
	title = {Toward lean satellites reliability improvement using {HORYU}-{IV} project as case study},
	volume = {133},
	issn = {0094-5765},
	url = {https://www.sciencedirect.com/science/article/pii/S009457651630724X},
	doi = {10.1016/j.actaastro.2016.12.030},
	pages = {33--49},
	journaltitle = {Acta Astronautica},
	shortjournal = {Acta Astronautica},
	author = {Faure, Pauline and Tanaka, Atomu and Cho, Mengu},
	urldate = {2021-05-20},
	date = {2017-04-01},
	langid = {english},
}

\end{document}